\documentclass[11pt,a4paper]{article}
\usepackage[hyperref]{template/tto2019}
\usepackage{times}
\usepackage{latexsym}
\usepackage[utf8]{inputenc} 
\usepackage[T1]{fontenc}    
\usepackage{hyperref}       
\usepackage{url}            
\usepackage{booktabs}       
\usepackage{amsfonts}       
\usepackage{nicefrac}       
\usepackage{microtype}      
\usepackage{lipsum}
\usepackage{graphicx}
\usepackage{amsmath}
\usepackage{color}
\usepackage{float}
\usepackage{url}
\usepackage{kotex}		    
\graphicspath{ {./images/} }

\hypersetup{
    colorlinks=true,
    linkcolor=blue,
    filecolor=cyan,      
    urlcolor=black,
}

\aclfinalcopy 

\makeatletter
\newcommand{\printfnsymbol}[1]{%
  \textsuperscript{\@fnsymbol{#1}}%
}
\makeatother
\definecolor{brown(web)}{rgb}{0.65, 0.16, 0.16}

\title{Risk Communication in Asian Countries: COVID-19 Discourse on Twitter}

\author{Sungkyu Park$^{1,2}\thanks{~~Equal contribution. Correspondence to: Meeyoung Cha <meeyoungcha@gmail.com>.}$~~~~~Sungwon Han$^{1,2}\printfnsymbol{1}$~~~~~Jeongwook Kim$^{1,2}\printfnsymbol{1}$~~~~~Mir Majid Molaie$^{1}$~~~~~Hoang Dieu Vu$^{3}$\\
\textbf{Karandeep Singh$^{2}$~~~~~Jiyoung Han$^{1}$~~~~~Wonjae Lee$^{1}$~~~~~Meeyoung Cha$^{2,1}$}
\\~\\
$^1$ Korea Advanced Institute of Science and Technology (KAIST), Daejeon, South Korea \\
$^2$ Institute for Basic Science (IBS), Daejeon, South Korea \\ 
$^3$ Hanoi University of Science and Technology (HUST), Hanoi, Vietnam}

\date{}

\begin{document}
\maketitle
\begin{abstract}
COVID-19 has become one of the most widely talked about topics on social media. This research characterizes risk communication patterns by analyzing the public discourse on the novel coronavirus from four Asian countries: South Korea, Iran, Vietnam, and India, which suffered the outbreak to different degrees. The temporal analysis shows that the official epidemic phases issued by governments do not match well with the online attention on COVID-19. This finding calls for a need to analyze the public discourse by new measures, such as topical dynamics. Here, we propose an automatic method to detect topical phase transitions and compare similarities in major topics across these countries over time. We examine the time lag difference between social media attention and confirmed patient counts.  
For dynamics, we find an inverse relationship between the tweet count and topical diversity.
%
\end{abstract}


\section{Introduction}


The novel coronavirus pandemic (COVID-19) has affected global health and the economy. Social media and Internet usage to seek and share information about the virus have increased massively~\cite{lazer2018science}, leaving them to be an excellent medium to examine the patterns of risk communication during a pandemic. Unfortunately, unconfirmed and intentional spread of false information can be seen on these platforms, jeopardizing public health. Studies have shown that people tend to share misinformation faster and more profoundly than real information~\cite{vosoughi2018spread,kwon2013prominent,kim2018leveraging}. The sheer amount of information and a mixture of right and wrong confuses people of what safety health guidelines to follow. A new term named Infodemic, which combines information and pandemic, has been newly introduced to describe this phenomenon\footnote{Coronavirus Disease 2019 (COVID-19) Situation Report. \href{https://bit.ly/2SKCl8X}{https://bit.ly/2SKCl8X}.}. In practice, Infodemic has already heavily impacted on society. For instance, the rhetoric of misinformation on COVID-19 has shifted from false preventive measures to the anti-vaccination movement~\cite{stecula2020trust} and vandalism towards telecommunication infrastructures~\cite{ahmed2020covid}.

In this research, we gathered data from social media to understand public discourse on the pandemic. Understanding the public concern will help determine which unproven claims or misinformation to debunk first, which contributes to fighting the disease. Primarily, we aim to discern what people say in the wild. For instance, identifying prevalent misinformation in a handful of countries first can help debunk the same piece of misinformation in other countries before the misinformation becomes a dominant topic and poses a threat to public health.

To detect timely topics by phase, we need to decide temporal phases first that can well reflect the real events and prevailing circumstances. If epidemic phases issued by each government are credible, we can directly use them. Otherwise, we can think of an alternative approach where we extract topics corresponding to the decided temporal phases by utilizing natural computational language processing methods. Based on those topics, we provide implications of the unique traits of risk communication locally and globally. This attempt helps to alleviate the propagation of false claims that can threaten public safety amid the COVID-19 pandemic. In this light, we have set up the following four research questions. Note that the developed codes can be accessed via Multimedia Appendix 1 and a GitHub page\footnote{The crawled Twitter dataset and the detailed information about the language-specific tokenizers is explained at \href{https://github.com/dscig/COVID19_tweetsTopic}{https://github.com/dscig/COVID19\_tweetsTopic.}}.

\begin{itemize}
\item
Can official epidemic phases issued by governments reflect the online interaction patterns?
\item
How to automatically divide topical phases based on a bottom-up approach?
\item
What are the major topics corresponding to each topical phase?
\item
What are the unique traits of the topical trends by country, and are there any notable online communicative characteristics that can be shared?
\end{itemize}

\section{Related Research}

\textbf{Issue Attention Cycle.} The issue attention cycle model can provide a pertinent theoretical framework for our analyses~\cite{downs1972issue}. The model conceptualizes how an issue rises into and fades away from the center of public attention. In the first stage, labeled as the pre-problem stage, an undesirable social condition (e.g., the appearance of COVID-19) emerges but has not yet captured much public attention. The second stage, alarming discovery and euphoric enthusiasm, occur when a triggering event (e.g., the national spike of newly confirmed cases of COVID-19 or WHO's a statement on COVID-19) heightens public awareness of the issue. In the third stage, realizing the cost of significant progress, people begin to recognize the hardship that requires a significant restructuring of society and significant sacrifices of some groups in the population to solve the problem. This causes a gradual decline of intense public interest, the fourth stage. In the final stage, the post-problem stage, the current issue is replaced by a new one and moves into a twilight zone of lesser public attention.

Not all issues follow the five stages of the issue attention cycle~\cite{nisbet2006attention}. As the cyclical patterns of public attention evolve, a wide array of public discourse has been found across multiple issues of climate change~\cite{mccomas1999telling}, emerging technologies~\cite{anderson2012news,wang2018framing}, and public health risks~\cite{shih2008media,arendt2019investigating}. There are also cultural differences~\cite{jung2012attention}. Despite these fragmented findings, issue attention cycle provides insights on how public attention dramatically waxes and wanes. An issue that has gone through the cycle is different from the issues that have not, with two respects, at least. First, amid the time that the issue earned the national prominence, new institutions, programs, and measures would have been developed to deal with the situation. These entities are likely to persist even after public attention has shifted elsewhere, thus having persistent societal impacts afterward. Second, these entities' prolonged impacts are subject to what was heavily discussed when the issue was the primary public concern.

With this regard, scholars need to look at the specifics of public conversations about a target issue. Although the issue attention cycle was initially conceived concerning traditional media, including newspapers and televisions, there is burgeoning literature applying the model to social media platforms. Most notably, in Twitter, the public is increasingly turning to for information seeking and sharing without the gate-keeping process~\cite{david2016tweeting}. Twitter conversations, as such, are more resonating with real-world word-of-mouth. It is not uncommon for journalists to refer to social media in their news stories. Research consistently finds that Twitter takes the initiative and greater control over public discourse, especially in the early stages of the issue-attention cycle~\cite{jang2017round,wang2018framing}. Building on these prior studies, we analyze the volume of Twitter conversations about COVID-19 to demonstrate an issue attention cycle on a social media platform.

%
\textbf{COVID-19-related Analyses.} Many studies have looked into the impact of the pandemic on various aspects. Some researchers have focused on predicting the transmissibility of the virus. One work estimated the viral reproduction number ($R_0$) of the virus. It showed $R_0$ of SARS-CoV-2 seems to be already more substantial than that of SARS-CoV, which was the cause of the SARS outbreak firstly found in the Guangdong province of China in 2002~\cite{liu2020reproductive}. Another work claims that by reducing 90\% of travel worldwide, the spread of epidemic could be significantly reduced, via a stochastic mathematical prediction model of the infection dynamics~\cite{chinazzi2020effect}.

Other lines of works are about understanding the propagation of misinformation related to COVID-19. One study modeled the spread of misinformation about COVID-19 as an epidemic model on various social media platforms like Twitter, Instagram, YouTube, Reddit, and Gab; it showed that users interact with each other differently and consume information differently depending on the platforms~\cite{cinelli2020covid}. In this light, media platforms like Facebook, YouTube, and Twitter claim to be trying to bring people back to a reliable source of medical information. To do so, they have direct communication lines with CDC and WHO~\cite{frenkel2020surge}.

When narrowing down to local-specific matters, one article claims that the fake news online in Japan has led to xenophobia towards patients and Chinese visitors, based on the qualitative analysis upon Japanese online news articles~\cite{shimizu20202019}. Meanwhile, a work surveyed with 300,000 online panel members in 2015, South Korea, while the MERS outbreak was prevalent in this country, and claimed that if the information from public health officials is untrustworthy, people rely more on online news outlets and communicate more via social media~\cite{jang2019information}. 

Another report argued that the public could not appreciate the information shared by public health officials due to prevalent misinformation on fake cures and conspiracy theories~\cite{oxfordmisinformation}. The efficacy of the response to restrain this Infodemic varies from country to country and depends on public confidence in the authorities. There is an attempt to compare three countries in terms of political bias. The authors conducted a large-scale survey across the US, the UK, and Canada. Statistically, they found that although political polarization of COVID-19 exists in the US and Canada, the exact belief in COVID-19 is broadly related to the quality of an individual's reasoning skills, regardless of political ideology~\cite{pennycook2020predictors}.

Many types of datasets are released to the public, as well as the research communities. One research crawled and opened tweet information from ten languages with the COVID-19-relevant keywords for around three months~\cite{chen2020covid}. Another work collated over 59K academic articles, including over 47K full research papers about COVID-19, SARS-CoV-2, and the related Coronavirus issues~\cite{wang2020cord} [27].

\section{Method}

\subsection{Data}

We crawled messages (tweets) from Twitter by using the Twint Python library\footnote{An advanced twitter scraping tool is written in Python. The detailed information about the scraper is explained at \href{https://github.com/twintproject/twint}{https://github.com/twintproject/twint.}} and the Search APIs\footnote{Official search tweets API for developers. Full-archive endpoint option provides complete access to tweets from the first tweet in March 2006. See also \href{https://developer.twitter.com/en/docs/tweets/search/}{https://developer.twitter.com/en/docs/tweets/search/}.}. Our focus is on South Korea, Iran, Vietnam, and India. These are all Asian countries, and thereby, we can control covariates concerning significant differences shown on social media between Western and Asian cultures~\cite{cho2013qualitative,li2018seeking}. These four countries were unique in terms of dealing with the current outbreak. In Iran, the number of confirmed cases has gradually increased since the first confirmed case, whereas in Vietnam, the numbers have consistently stayed (relatively) low. There was an abrupt increase in the numbers after the first confirmed case in Korea, but it seems they have successfully flattened the rising curve of confirmed cases, unlike other countries. In India, the situation had been relatively mild until mid-Mar, since then, there has been a drastic surge.

We have set up two keywords, ``Corona'' and ``Wuhan pneumonia'' to crawl tweets (see Table~\ref{table:stat_tweets} to find exact keywords used for crawling tweets for each country) and collected tweets for the three months from January to March in 2020.

\begin{table}[t!]
\small
\centering
\begin{tabular}{llll}
\toprule
\textbf{Language} & \textbf{Duration} & \textbf{Used Keyword$^\dag$} & \textbf{\# of Tweets} \\ 
\midrule
Korean     & Jan 1 to     & Corona,         & 1,447,489 \\ 
           & Mar 27, 2020 & Wuhan pneumonia & \\           
Farsi      & Jan 1 to     & \#Corona,       & 459,610 \\   
           & Mar 30, 2020 & \#Coronavirus,  & \\           
           &              & \#Wuhan,        & \\           
           &              & \#pneumonia     & \\           
Vietnamese & Jan 1 to     & corona,         & 87,763 \\    
           & Mar 31, 2020 & n-cov,          & \\           
           &              & covid,          & \\           
           &              & acute pneumonia & \\           
Hindi      & Jan 1 to     & Corona,         & 1,373,333 \\ 
           & Mar 31, 2020 & Wuhan pneumonia & \\           
\bottomrule
\multicolumn{4}{l}{\small{$^\dag$ Keywords are listed here after translated in English from the}} \\
\multicolumn{4}{l}{\small{actual local languages, e.g., ``코로나'' $\xrightarrow{}$ ``Corona'' in Korean.}}
\end{tabular}
\caption{Statistics of the crawled tweets.}
\label{table:stat_tweets}
\end{table}

\subsection{Pipeline for Detecting Topical Phases then Extracting Topics}

\begin{figure}[t]
\centerline{\includegraphics[width=1.05\linewidth]{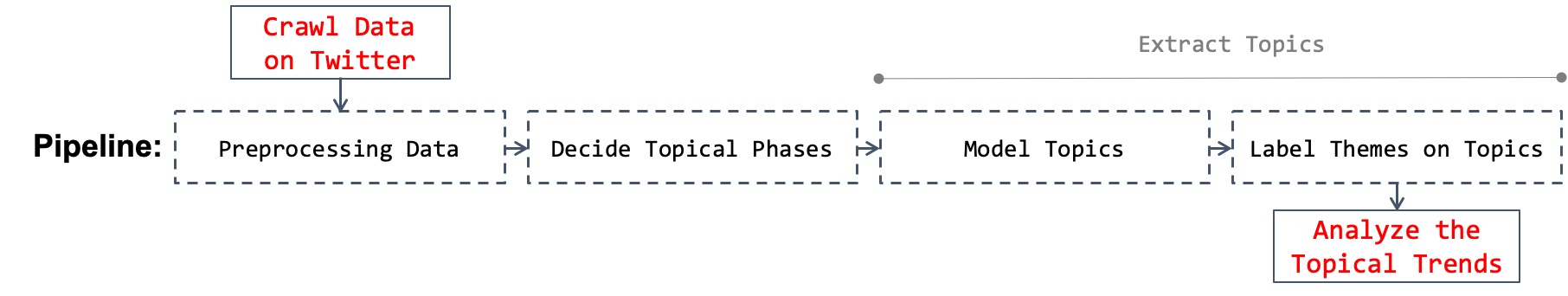}}
\caption{The pipeline of the topic analysis.}
\label{fig:pipeline.designs}
\end{figure}

The data collection pipeline includes the following four modules to eventually extract and label major topics for certain phases, as shown in Figure~\ref{fig:pipeline.designs}. This process is repeated for all four countries. \\

\textbf{Preprocessing Data.} To extract topics from the collected tweets by NLP, we first need to tokenize the data, which can be defined as converting data to the smallest units that have meaning. We have filtered unnecessary textual information like stop words, special characters (non-letters), special commands, and emojis. We then utilize the existing Python tokenizer libraries corresponding to each specific language. Detailed information about the language-specific tokenizers is explained at the provided web link. \\

\textbf{Decide Topical Phases.} The next step is to set up specific target phases divided by dates to extract topics. This is nontrivial since there are multiple fluctuations and changes on the topics reflecting the real events such as an increase in COVID-19 patients. Furthermore, we do not consider to use the epidemic phases announced by governments since the offline epidemic phases seem not to capture the actual online topic trends, as will be explained at the forthcoming Basic Daily Trends section.

We, therefore, devise a bottom-up approach to decide dates that show the sign of sudden increases in the daily volume of the tweets. We set up two learnable parameters of the first derivatives (hereafter $velocity$) and the second derivatives (hereafter $acceleration$) of the daily tweet volumes, as illustrated in the formulas below where $D$ is a day, $t$ is a target date, and $t-1$ is one past date from $t$.

\begin{equation}
\begin{split}
\textrm{\textit{velocity}} = \frac{\textrm{\textit{\# of $tweet_t$}}-\textrm{\textit{\# of $tweet_{t-1}$}}}{ \textrm{\textit{$D_t$}}-\textrm{\textit{$D_{t-1}$}}} \\
\textrm{\textit{acceleration}} = \frac{ \textrm{\textit{$velocity_t$}}-\textrm{\textit{$velocity_{t-1}$}}}{ \textrm{\textit{$D_t$}}-\textrm{\textit{$D_{t-1}$}}}
\end{split}
\end{equation}

We reckoned the $velocity$ and $acceleration$ values when the first confirmed case was announced as the ground truths (GT) by country. The intuition of this approach is that the $velocity$ and $acceleration$ values are proxies to unique communication traits for each country in terms of a specific subject (i.e., COVID-19 in our case). Once they have been computed from the first confirmed date, they would be the same for the following period.

We have set up joint thresholds for $velocity$ and $acceleration$ to find dates that show while $velocity$ is still smaller than the $velocity_{GT}$, $acceleration$ becomes more substantial than the $acceleration_{GT}$: 0 < $velocity$ < $velocity_{GT}$ \& $acceleration$ > $acceleration_{GT}$. In this light, we learn two parameters from the first confirmed date by country then detect other dates that can be conjectured as the start of forthcoming topical phases. When learning parameters, for $velocity$, we round down the $velocity_{GT}$ value and add 1. For $acceleration$, we round down the $acceleration_{GT}$, which is similar to the loss minimization concept of the machine-learning approach (i.e., a learning process is finished by one step).

We adopted a low-pass filter with 0.2 as the low-frequency threshold to remove noisy signals to smoothen the data. Finally, the temporal data was divided into topical phases (See Appendix 1 to find the computed daily velocity and acceleration trends and decided phases accordingly by country). \\

\textbf{Extract Topics -- Model Topics.} We have utilized the latent Dirichlet allocation (LDA) for the topic modeling task. The LDA is a well-known machine-learning method to extract topics amid given textual documents (i.e., a collection of discrete data-points) – tweets in our case~\cite{ostrowski2015using}. The LDA generates and maximizes the joint probability between the word distribution of topics and the topic distribution of documents~\cite{blei2003latent}. The number of topics for each phase is a hyperparameter. We have set the range of the number of topics is between 2 and 50, and calculate perplexity (PPL), probability of how many tokens can be placed at the next step (i.e., indicating the ambiguity of the possible next token). PPL is a well-known metric to optimize a language model with a training practice~\cite{adiwardana2020towards}. During the iteration, we have fixed the minimum required frequency of words among the entire tweets for each phase to be 20 and the epoch number for each topic to be 100, respectively. We then decide the optimum number of topics for each phase by choosing the minimum PPLs.

As a result, we have decided on the number of topical phases and the corresponding optimized number of topics for each phase, as presented in Table~\ref{table:optimzed_no_topics}. For example, in the case of South Korea, after the number of phases was decided as four from the `Decide Topical Phases' module, the optimized number of topics is computed for each phase as 2, 41, 15, and 43, respectively. \\

\begin{table*}[t!]
\small
\centering
\begin{tabular}{lcccccc}
\toprule
\textbf{Country} & \textbf{Phase 0} & \textbf{Phase 1} & \textbf{Phase 2} & \textbf{Phase 3} & \textbf{Phase 4} & \textbf{Phase 5} \\
\midrule
South Korea                   & Jan 1-19 & Jan 20-Feb 12 & Feb 13-Mar 9 & Mar 10-27 & - & - \\
\small{\textit{velocity}: 274} \tiny{($tweets/day$)} & 14.06  & 2,415.52 & 5,376.769 & 5,577.88  &  & \\
\small{\textit{acceleration}: 109} \tiny{($t^{2}/d$)}& 28.17  & 5,244.09 & 17,796.08 & 13,095.65 &  & \\
                              & 21.78  & 56,809.78    & 211,310.89 & 147,759.41 &  & \\
                              & 0.77  & 10.83   & \textbf{\texttt{11.87}} & 11.28 &  & \\
                              & 2$\xrightarrow{}$1$\xrightarrow{}$1 & 41$\xrightarrow{}$18$\xrightarrow{}$8 & 15$\xrightarrow{}$6$\xrightarrow{}$5 & 43$\xrightarrow{}$21$\xrightarrow{}$11 &  &  \\
Iran                          & Jan 1-Feb 18 & Feb 19-Mar 30 & - & - & - & - \\
\small{\textit{vel}: 1,724}   & 245.34  & 1,442.46   &  &  &  & \\
\small{\textit{acc}: 787}     & 385.63  & 5,272.04 &  &  &  & \\
                              & 1,315.13  & 22,128.76   &  &  &  & \\
                              & 3.41  & \textbf{\texttt{4.20}}   &  &  &  & \\                              
                              & 3$\xrightarrow{}$3$\xrightarrow{}$3 & 5$\xrightarrow{}$4$\xrightarrow{}$6 &  &  &  & \\
Vietnam                       & Jan 1-20 & Jan 21-25 & Jan 26-Feb 15 & Feb 16-Mar 4 & Mar 5-22 & Mar 23-31 \\
\small{\textit{vel}: 49}      & 3.79  & 131.25    & 179.65 & 485.59 & 340.65 & 433.29 \\
\small{\textit{acc}: 23}      & 7.37  & 218.50    & 686.60 & 1,238.77 & 1,089.94 & 1,224.00 \\
                              & 0.21  & 20.75    & 159.80 & 582.29 & 192.24 & 201.86 \\
                              & 0.03  & 0.09    & 0.23 & \textbf{\texttt{0.47}} & 0.18 & 0.16 \\
                              & 19$\xrightarrow{}$1$\xrightarrow{}$1 & 3$\xrightarrow{}$1$\xrightarrow{}$2 & 6$\xrightarrow{}$3$\xrightarrow{}$4 & 46$\xrightarrow{}$22$\xrightarrow{}$7 & 48$\xrightarrow{}$20$\xrightarrow{}$10 & 16$\xrightarrow{}$4$\xrightarrow{}$2 \\
India                         & Jan 1-29 & Jan 30-Mar 9 & Mar 10-Mar 31 & - & - & - \\
\small{\textit{vel}: 783}     & 107.41 & 1,364.95 & 13,318.63 &  &  & \\
\small{\textit{acc}: 285}     & 269.72 & 4,261.13 & 58,924.55 &  & \\
                              & 415.69 & 14,467.8 & 318,368.05 &  &  & \\
                              & 1.54 &  3.40 & \textbf{\texttt{5.40}} &  &  & \\                              
                              & 3$\xrightarrow{}$1$\xrightarrow{}$3 & 50$\xrightarrow{}$22$\xrightarrow{}$5 & 47$\xrightarrow{}$22$\xrightarrow{}$9 &  &  &  \\
\bottomrule
\end{tabular}
\caption{The extracted \# of phases by country and the optimized \# of topics within. First row by country: Time Period; Second: \# of Users per Day; Third: \# of Tweets per Day (\textit{A}); Fourth: \# of Retweets per Day (\textit{B}); Fifth: Tweet Depth (\textit{B/A}); Sixth: Optimized \# of Topics based on PPL$\xrightarrow{}$Major (i.e., 75\% percentile) Topics$\xrightarrow{}$Final \# of Merged Theme Labels from Human Annotators.}
\label{table:optimzed_no_topics}
\end{table*}

\textbf{Extract Topics -- Label Topics.} This step involves labeling the main themes for the extracted topics. This is to allocate semantic meanings to each topic and to analyze the semantic trends. We first sorted all tweets with the estimated topic numbering by descending order (i.e., tweets with larger volumes in terms of the estimated topic numbering list first) and discarded the minor topics that accounted for less than 25\% percentile of all tweets.

We then extract 1K most retweeted tweets and 30 highest probable keywords for each topic. We provide these datasets to domain experts from each country and ask them to label themes for each topic based on the given datasets. Via qualitative coding, any similar or hierarchical topics were merged into a higher category. The final count of themes is shown in the third row for each country in Table~\ref{table:optimzed_no_topics}. Besides, if one topic corresponds to several themes, then it is labeled to have multiple classes. The maximum number of multiple cases within topics was two, and each case within a topic was weighed as 0.5 when plotting the daily trends of the tweet counts.

Concerning the local/global news themes, we have narrowed down the labels since people talk about different issues under the news category. We have sub-labeled them as \textit{\_confirmed} if tweets are about the confirmed/death cases, \textit{\_hate} if about the hate crimes towards individual races, \textit{\_economy} if about the economic situations/policies, \textit{\_cheerup} if about supporting each other, \textit{\_education} if about when to reopen schools, and none if about general information, respectively.

\section{Result}


\subsection{Basic Daily Trends}
\begin{figure}[!ht]
\centerline{\includegraphics[width=1.10\linewidth]{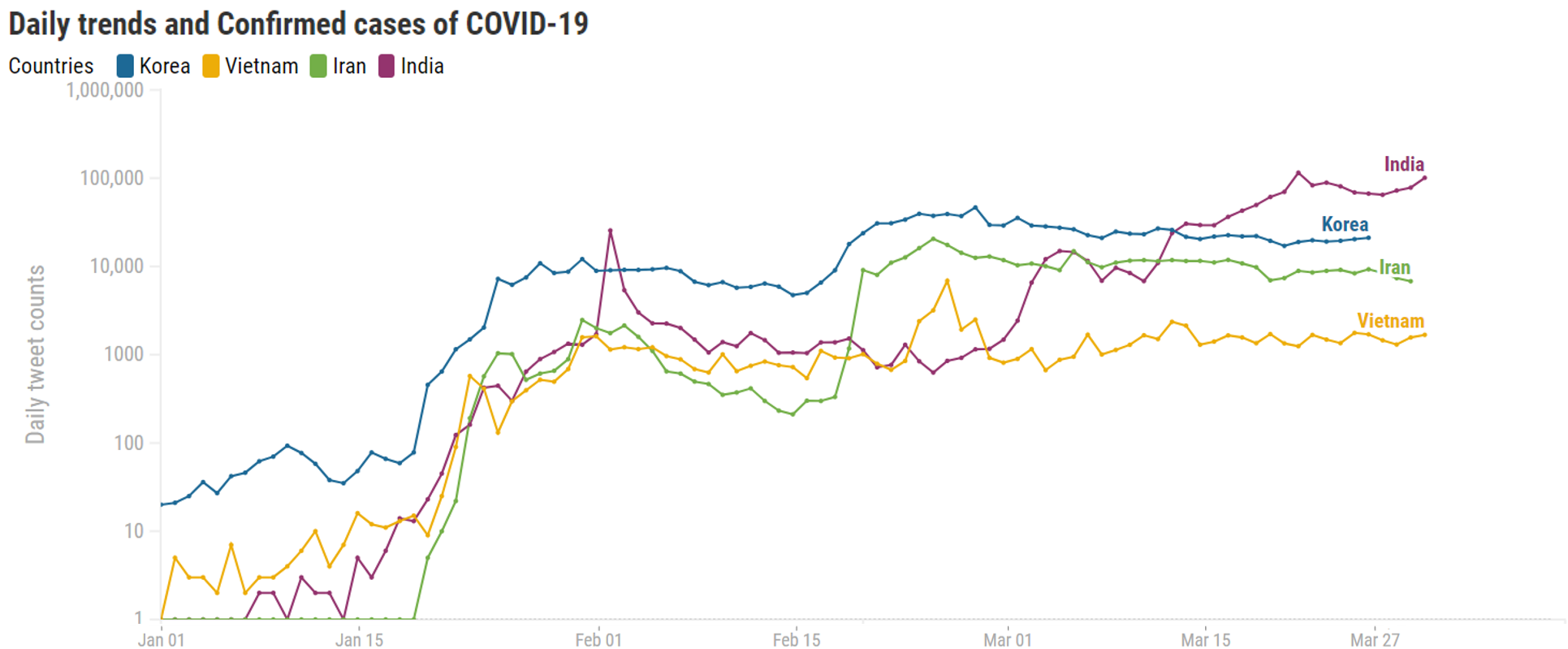}}
\caption{Daily trends on the Four countries: X-axis is dates and Y-axis is trends of \# of tweets with log-scale.}
\label{fig:trends.overall}
\end{figure}

\begin{figure}[!ht]
\centerline{\includegraphics[width=1.05\linewidth]{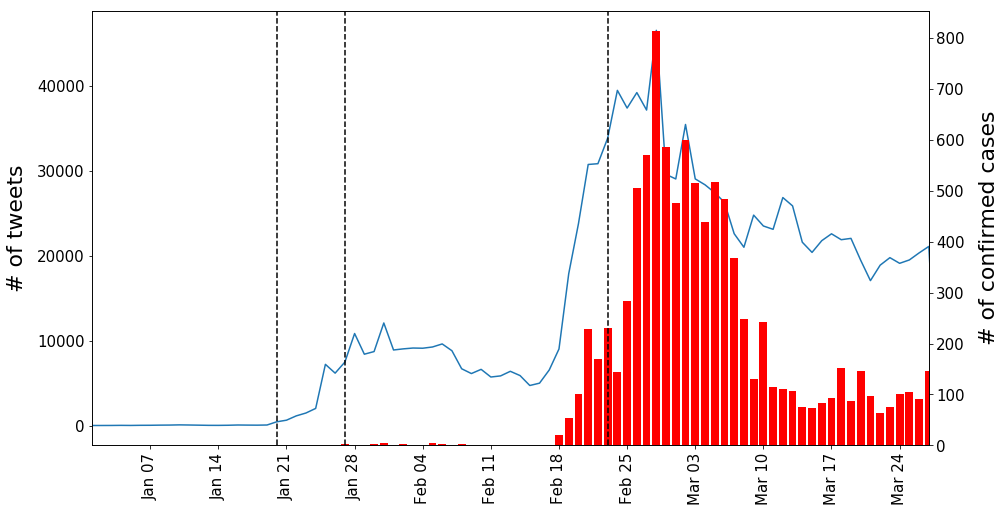}}
\caption{Daily trends on South Korea: start/end dates of the official epidemic phases (vertical dash lines), trends of \# of tweets (blue lines), and that of \# of the confirmed cases (red bars).}
\label{fig:trends.korea}
\end{figure}

\begin{figure}[!ht]
\centerline{\includegraphics[width=1.05\linewidth]{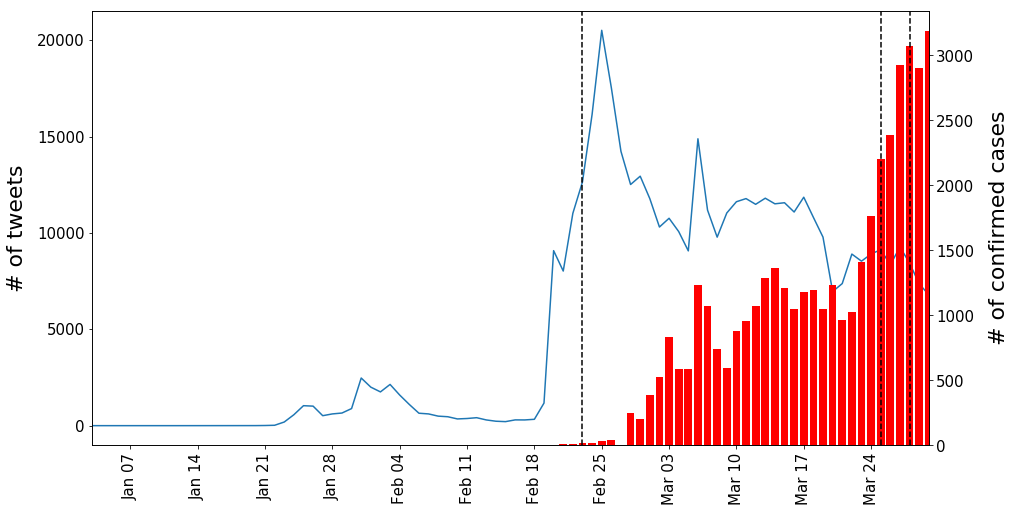}}
\caption{Daily trends on Iran: start/end dates of the official epidemic phases (vertical dash lines), trends of \# of tweets (blue lines), and that of \# of the confirmed cases (red bars).}
\label{fig:trends.iran}
\end{figure}

\begin{figure}[!ht]
\centerline{\includegraphics[width=1.05\linewidth]{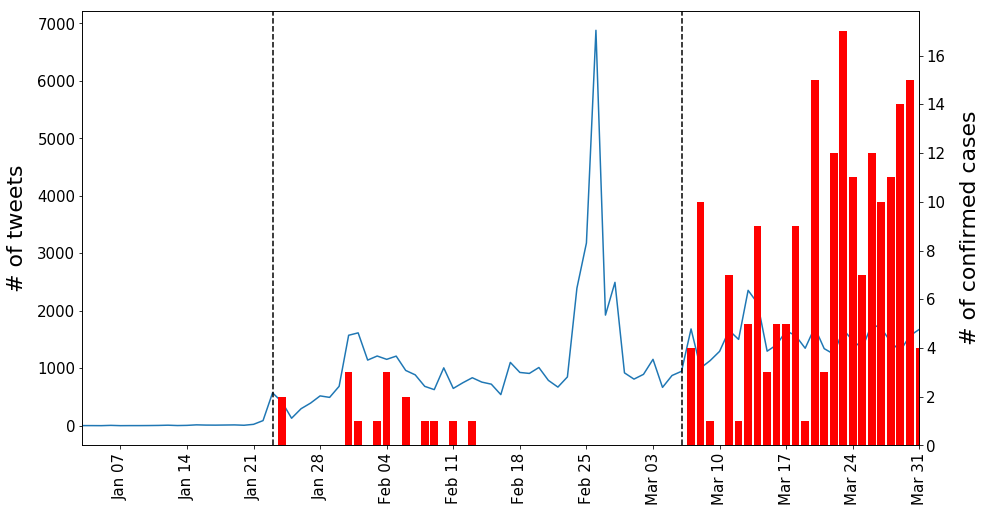}}
\caption{Daily trends on Vietnam: start/end dates of the official epidemic phases (vertical dash lines), trends of \# of tweets (blue lines), and that of \# of the confirmed cases (red bars).}
\label{fig:trends.vietnam}
\end{figure}

\begin{figure}[!ht]
\centerline{\includegraphics[width=1.05\linewidth]{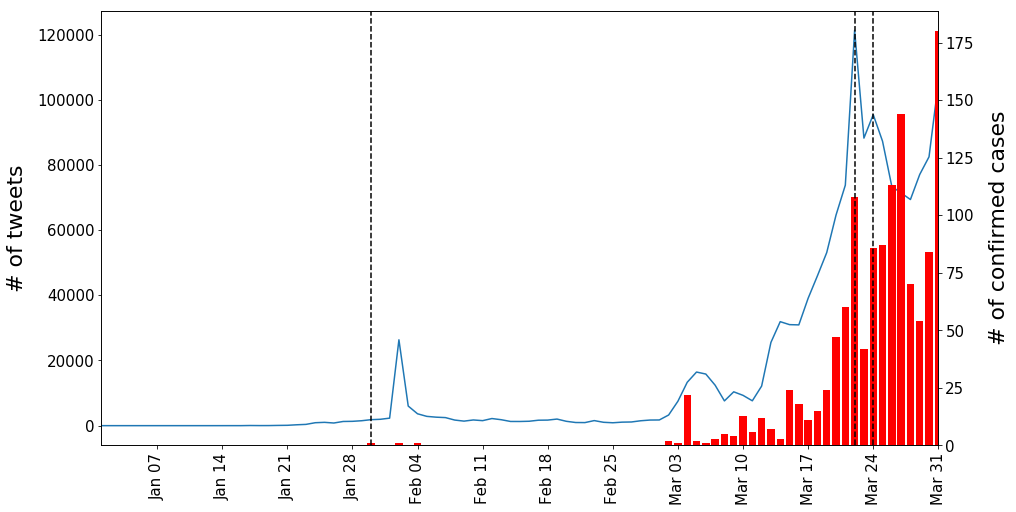}}
\caption{Daily trends on India: start/end dates of the official epidemic phases (vertical dash lines), trends of \# of tweets (blue lines), and that of \# of the confirmed cases (red bars).}
\label{fig:trends.india}
\end{figure}

We first depict the trends by plotting the daily number of tweets (see Figure~\ref{fig:trends.overall}). We then see the daily tweet counts, and the daily number of the COVID-19 confirmed cases simultaneously by country, as depicted in Figure~\ref{fig:trends.korea}--\ref{fig:trends.india}: adding to the two trends, we include official epidemic phases announced by each government as vertical lines. We confirmed that the tweet trends are associated with the confirmed case trends by seeing the tweet and confirmed case trends. Yet, the epidemic phases do not explain the tweet trends accurately. \\

\textbf{South Korea.} The first COVID-19 patient was reported on January 20, 2020\footnote{ COVID-19 pandemic in Korea. Wikipedia 2020. URL: https://bit.ly/3fy4SZp.}. This explains why the tweet count remains relatively low during early January, and it increases mainly only after late January (see Figure~\ref{fig:trends.korea}). On January 25, the Korean government issued a travel warning on Wuhan and the Hubei province, as well as the suggested evacuation of Korean citizens from those areas, which was primarily discussed on Twitter. On February 18, the numbers increased sharply due to the 31st confirmed case, which was related to a cult religious group "Shincheonji" in Daegu city. After the 31st confirmed case had been found, the quarantine authority tried rigorous testing focusing on Daegu, and the number of the confirmed cases were drastically increasing until mid-March. The tweet trends follow the same pattern. However, the official epidemic phases announced by the government, divided by the vertical dash lines in the figure, seem to lag from the increasing number of tweets. This pattern shows that the official epidemic phases do not match well with online attention.

\textbf{Other countries.} We have repeated the same analysis with three other countries, as shown in Figure~\ref{fig:trends.iran}--\ref{fig:trends.india} (See Appendix 2 to find the country's detailed explanations).

\subsection{Extracted Topical Trends}
We utilized the theme labels acquired from the `Label Topics' module daily and analyzed the topic changes across time with plots for the four target countries. One plot shows daily trends based on the number of tweets. In contrast, another plot shows the trends based on the number of tweets that contained mentions of country names like the U.S. Overall, as people increasingly talk more on the COVID-19 outbreak (i.e., the daily number of tweets increase), the topics people talk about become less diverse. \\

\begin{figure}[hbt!]
\centerline{\includegraphics[width=0.95\linewidth]{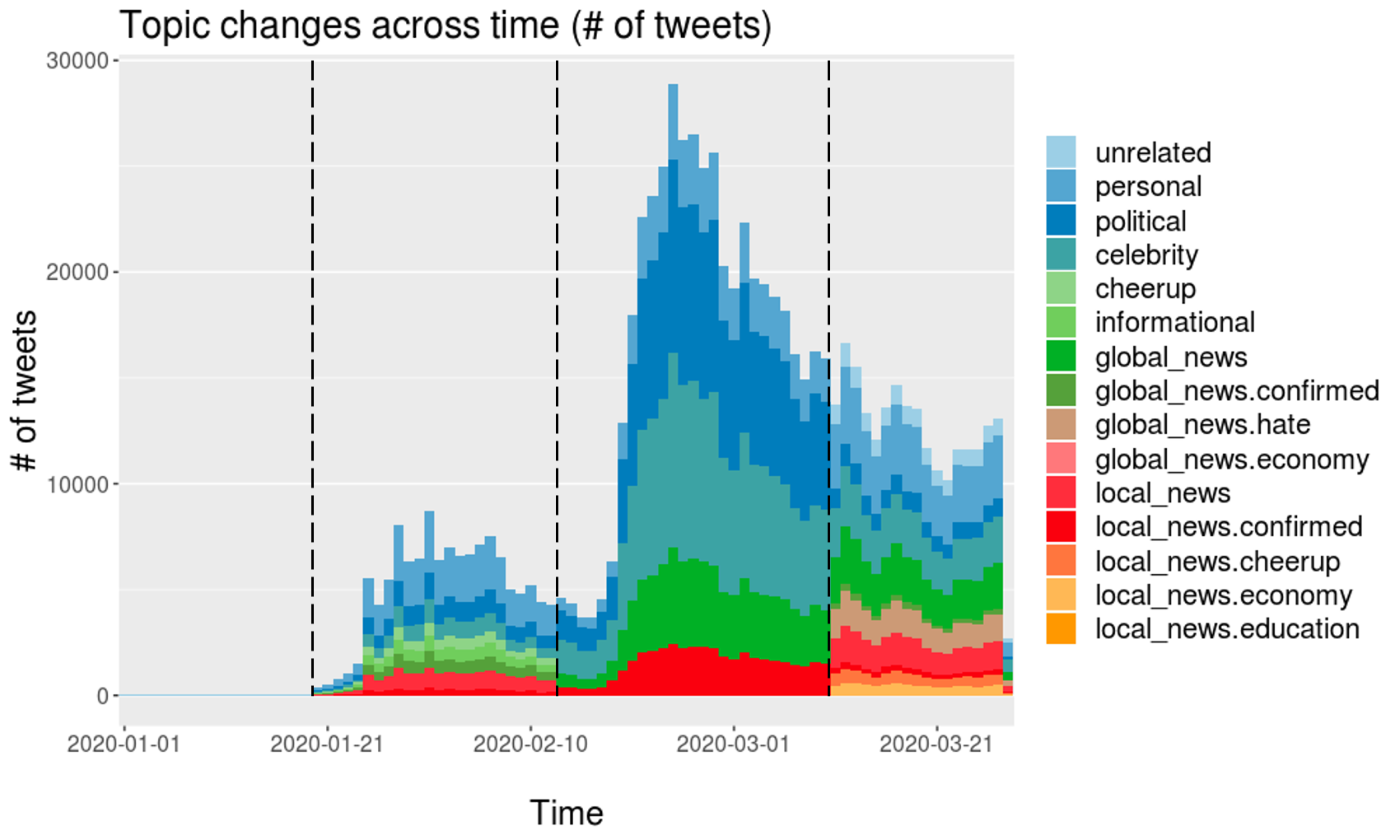}}
\centerline{\includegraphics[width=0.95\linewidth]{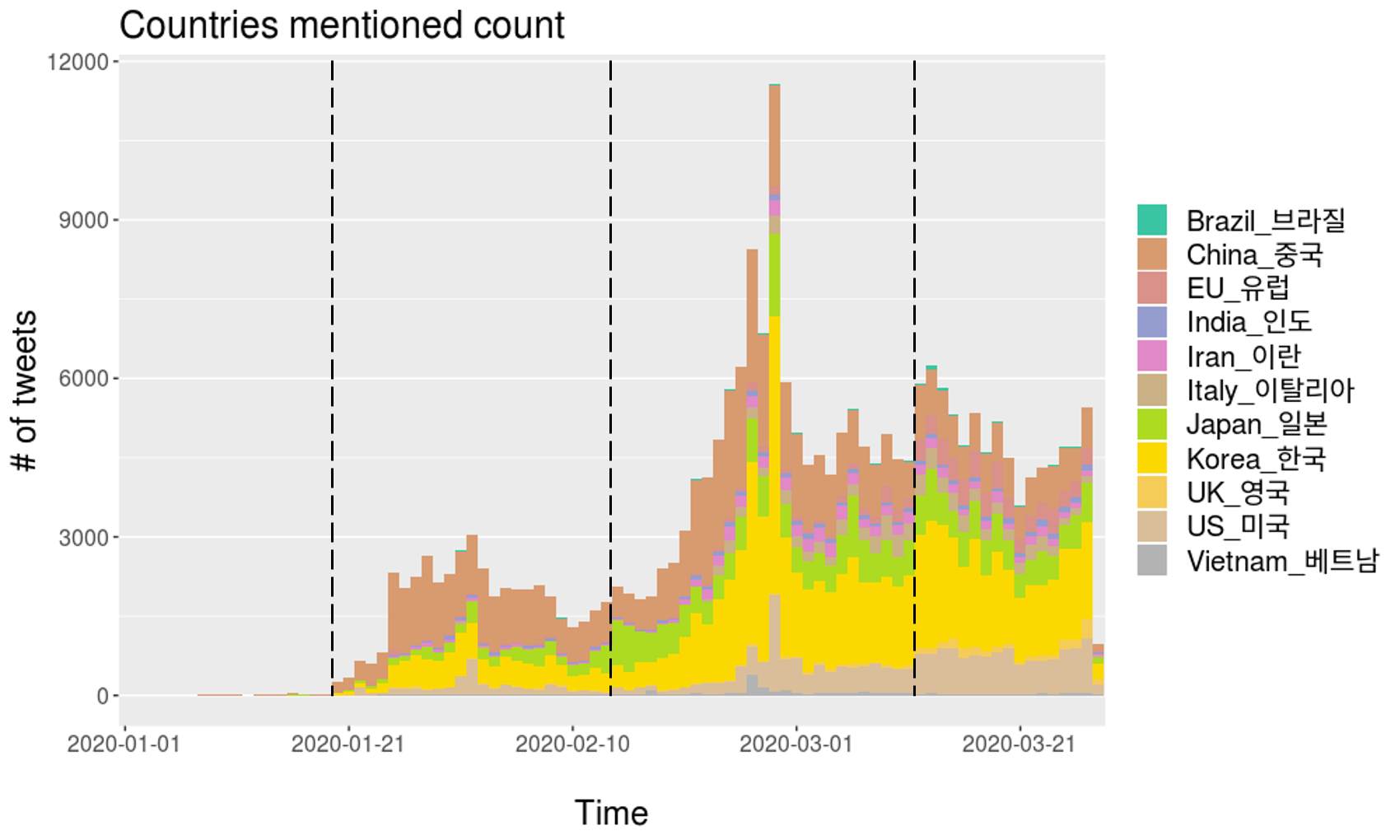}}
\caption{Daily topical trends on South Korea: based on \# of tweets (top) and based on \# of tweets country names mentioned (bottom).}
\label{fig:topic_change_korea}
\end{figure}

\textbf{South Korea.} The data yielded a total of four topical phases, which are used in Figure~\ref{fig:topic_change_korea}. Phase 0 has no related topic. Then from Phase 1 to Phase 3, the number of topics diverged as 8, 5, and 11. In Phase 1, people talk much about personal thoughts and opinions linked to the current outbreak, and they cheered each other. On Phase 2, as the crisis going up to its peak, people talked less on personal issues and mainly talked about political and celebrity issues. The political issues were about shutting down the borders of South Korea for China and other countries for Korea. On Phase 3, as the daily number of tweets becomes smaller than Phase 2, people tended to talk on more diverse topics, including local and global news. In particular, people were worried about hate crimes directed towards Asians in Western countries. People might be interested in different subjects as they think the crisis seems to be off the peak.

We see the daily trends on the mention of other countries by counting the tweets remarking on other country's name, either in their local languages or in English. Korea, China, and Japan were mostly mentioned, and we suspect that political and diplomatic relationships mainly triggered it. Meanwhile, the US and Italy similarly were steadily mentioned across the three months, and the media outlets broadcasting global news affect this phenomenon.

\textbf{Other countries.} We repeat the same analysis and interpret the results for other cases, Iran, Vietnam, and India, as depicted in Figure~\ref{fig:topic_change_iran}--\ref{fig:topic_change_india}. See Appendix 3 for the derived topical trend graphs and the detailed corresponding explanations by country.

\begin{figure}[hbt!]
\centerline{\includegraphics[width=1.05\linewidth]{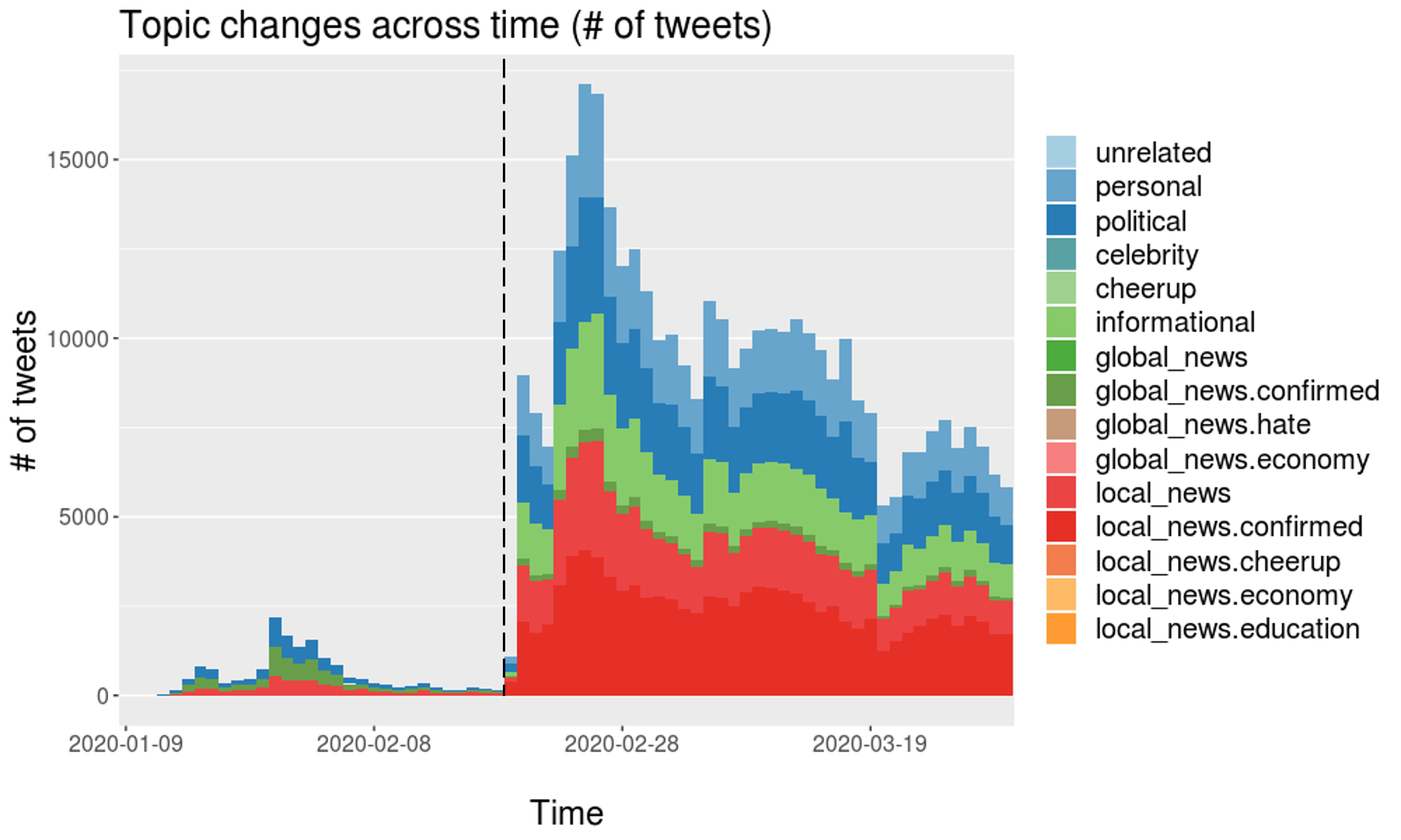}}
\caption{Daily topical trends on Iran: based on \# of tweets.}
\label{fig:topic_change_iran}
\end{figure}

\begin{figure}[hbt!]
\centerline{\includegraphics[width=1.05\linewidth]{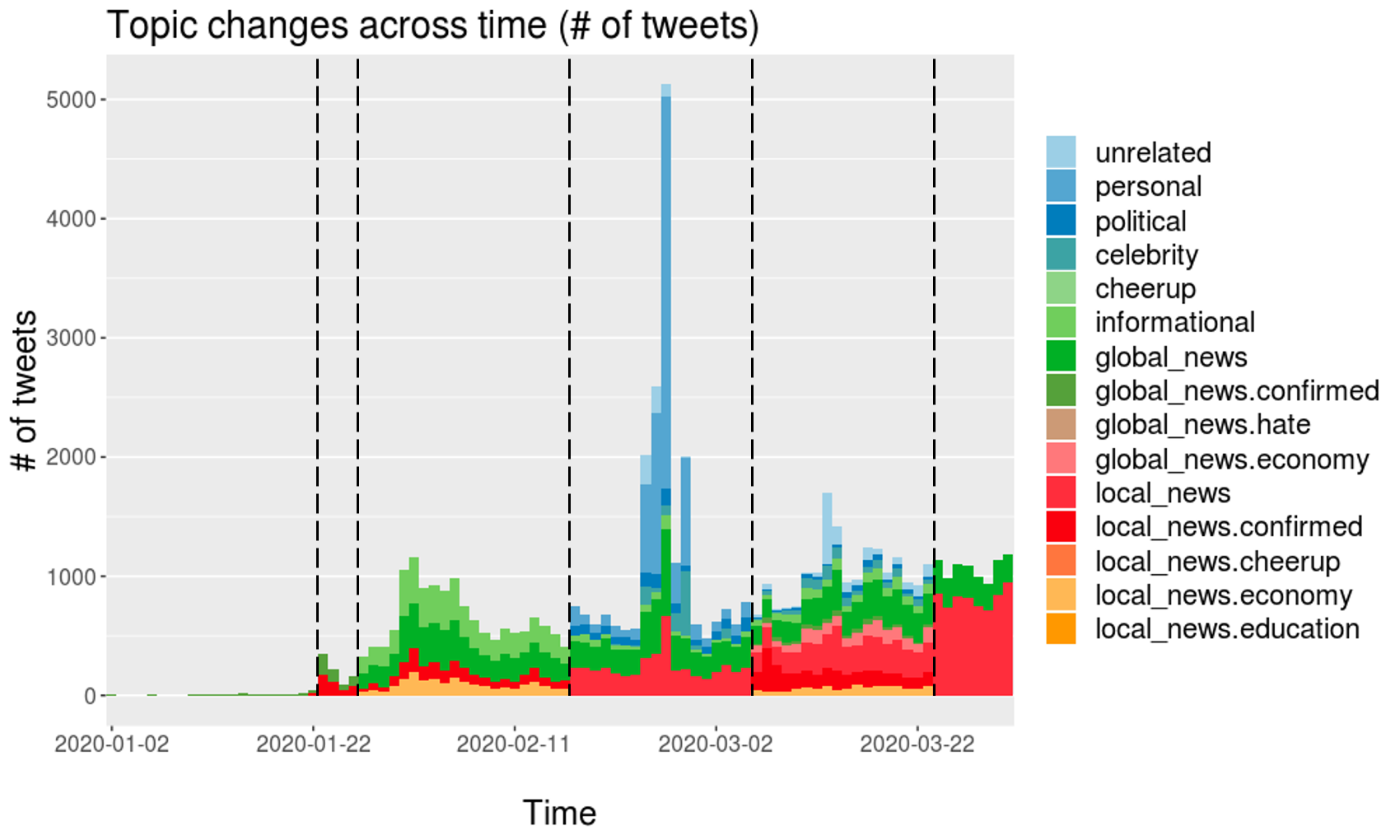}}
\caption{Daily topical trends on Vietnam: based on \# of tweets.}
\label{fig:topic_change_vietnam}
\end{figure}

\begin{figure}[hbt!]
\centerline{\includegraphics[width=1.05\linewidth]{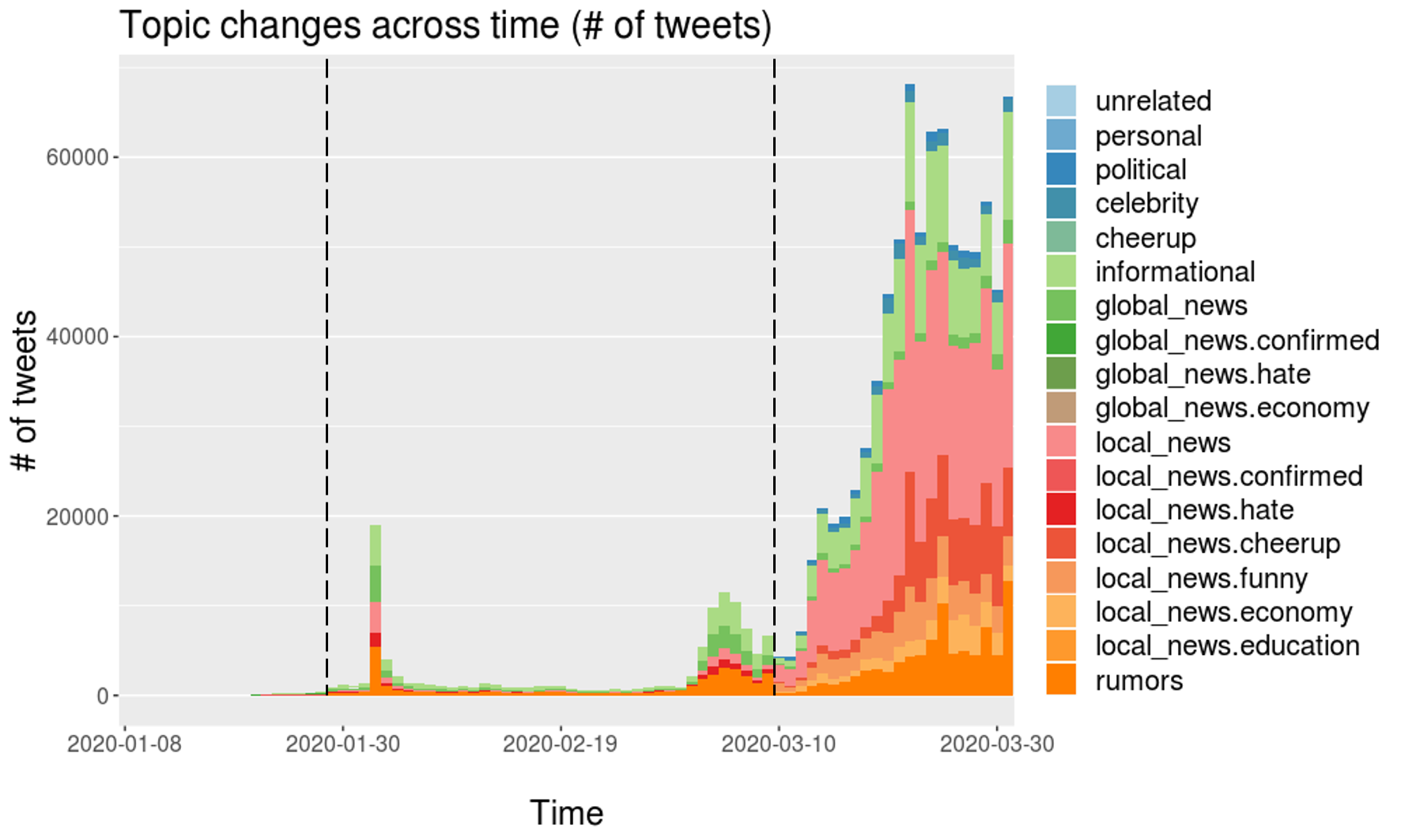}}
\caption{Daily topical trends on India: based on \# of tweets.}
\label{fig:topic_change_india}
\end{figure}

\section{Discussion}

We have analyzed tweets to understand the public discourse on the COVID-19 pandemic. In South Korea, the daily numbers of tweets reached local maxima with major offline events. However, in the case of Iran and Vietnam, the tweet counts did not synchronize well with offline events. It may be because Twitter is not as widely used in these two countries. Overall, it is interesting to observe that peaks in Twitter data do not necessarily correlate with significant events identified by local governments. Therefore, we use a bottom-up approach to explore the topical phrases which resonate with the flow of the open views.

After extracting the topical phases, number 4 in South Korea, 2 in Iran, 6 in Vietnam, and 3 in India, respectively, we used the LDA, and found the optimum number of topics for each topical phase, and then labeled the corresponding topics with appropriate themes. In general, as people talk more about COVID-19, the topics they refer to tend to be smaller in number. This was more apparent when tweet depth value is used for phases, as presented in Table~\ref{table:optimzed_no_topics}. Tweet depth is defined as the number of retweets per day divided by the number of tweets per day. It can be deemed as a standardized cascading depth, and therefore, the more considerable value signifies the greater extent of the depth for one tweet. From the case from South Korea and Vietnam, we could verify the observation as tweet depth tends to get larger when people communicate more on COVID-19. This phenomenon reaffirms another research that found the diameter value of the online Coronavirus network was lower than that of others~\cite{park2020conversations}. However, for Iran and India, the number of phases were too small to observe the topical trends' general traits.

Moreover, we found that the daily tweet trend peak succeeded in the daily confirmed cases. In Iran, Vietnam, and India, the peak of the daily tweet trend preceded the peak of the daily confirmed cases up to a few weeks. Although the two peaks are close to each other in South Korea, it is worth noting that around that time, the country was becoming the most affected by COVID-19 outside mainland China. Interestingly, as shown in Figure 3-6, the upsurge of the number of tweets in South Korea, Iran, and Vietnam (except India) was simultaneously observed at the end of February, before the upsurge of the locally confirmed cases. Given that COVID-19 is a global issue, this suggests that the issue attention cycle of COVID-19 on a social media platform is more responsive to global events than local ones. In this regard, the COVID-19 pandemic offers an exciting opportunity for future research to theorize the issue attention cycle model on a global scale and see how it evolves in conjunction with local specific topics such as increasing or decreasing confirmed cases, government measures, and social conflicts.

When comparing South Korea and Vietnam, there is an intriguing point to discuss. The topic of Phase 0 in Korea was not related to COVID-19, whereas Vietnam was about global news with confirmed cases. We do not attempt to generalize any findings due to the small tweet volume in Phase 0 for both countries. Still, Vietnamese users discussed the global epidemic issue more from the first place, and this tendency affects successful defending against pandemic later on.

To be specific to each country, in case of South Korea, when the local pandemic (offline) situation has become severe (Phase 2), the number of topics becomes smaller, which means people focus more on a handful of issues. A unique trait can be observed that in the (phase 0), people cheered each other up and hustled to express solidarity in the difficult times. In the case of Iran, the number of topics has been relatively steady across time, while the significant topics discussed have been confined to news and information: we interpret that Iranian users tend to be cautious about using social media. In case of Vietnam, at Phase 4, where the tweet traffic is relatively lower as compared to the Phase 3, the number of topics becomes more substantial, and the themes of topics become less direct to the confirmed and death tolls, e.g., people talked about the economy in Phase 2 and 4. Meanwhile, the Indian case indicates another unique trait: many topics were mainly related to misinformation, the scale of which was much lower in other countries. A large portion of misinformation, disinformation, and hateful contents is steadily observed on both Phase 1 and 2 (see Multimedia Appendix 3).

\section{Concluding Remark}

There are several limitations to be considered. First, we analyzed tweets solely from the four countries, and therefore, we need to be cautious about addressing explanations and insights that can be applied in general. We plan to extend the current study by including more countries. Second, there are other ways to decide the topical phases. Our approach can be aligned with the issue attention cycle as we compute unique communication traits (i.e., $velocity$ and $acceleration$ by country) that would be relatively consistent in a pandemic of COVID-19.

Nonetheless, the current research provides a valuable picture of critical topics from multiple countries on COVID-19. We automatically divide topical phases and extract major topics by phase. We then find several issues that were uniquely manifested in the recent pandemic crisis by each country. For instance, we may discover the emergence of misinformation in Hindi tweets. Our findings shed light on understanding public concerns and misconceptions under the crisis and, therefore, can help determine what misinformation is to be discredited with priority. This attempt helps defeat the Infodemic and limit the spread of the pandemic.
\bibliography{_reference}
\bibliographystyle{template/acl_natbib}
\clearpage
\section{Appendices}
\subsection{Appendix 1. Computed Daily Velocity/Acceleration Trends and Decided Temporal Phases by Country}

\textbf{\\}
\subsubsection{South Korea}

\begin{figure}[h!]
\centerline{\includegraphics[width=1.35\linewidth]{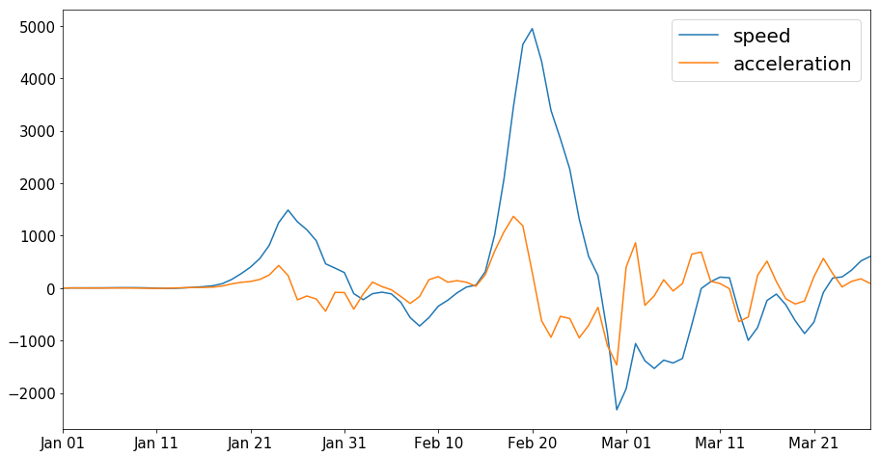}}
\textbf{\\}
\centerline{\includegraphics[width=1.35\linewidth]{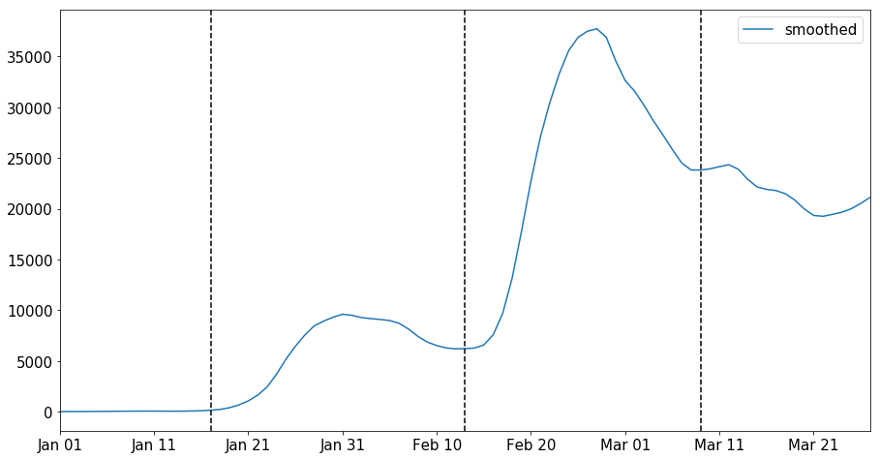}}
\caption{The South Korean case: Daily trends on velocity and acceleration of the \# of tweets (top) and divided phases detected by vertical dash lines (bottom).}
\label{fig:vel_acc_korea}
\end{figure}

\clearpage
\textbf{\\ \\}
\subsubsection{Iran}

\begin{figure}[h!]
\centerline{\includegraphics[width=1.35\linewidth]{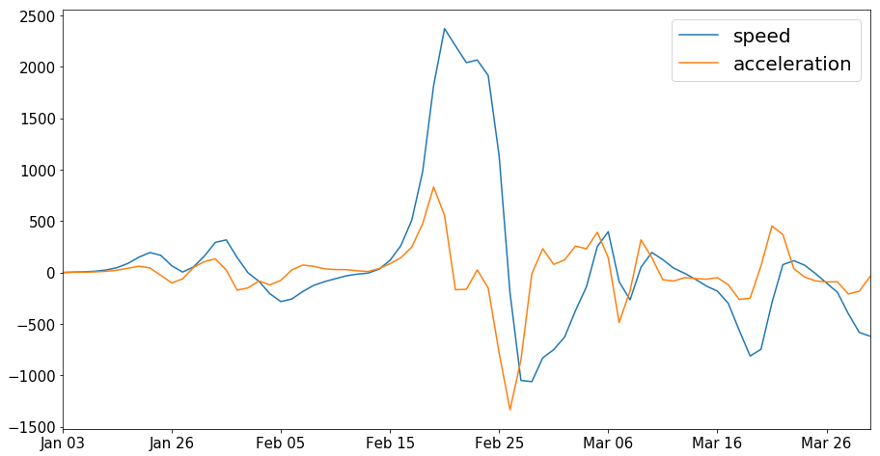}}
\textbf{\\}
\centerline{\includegraphics[width=1.35\linewidth]{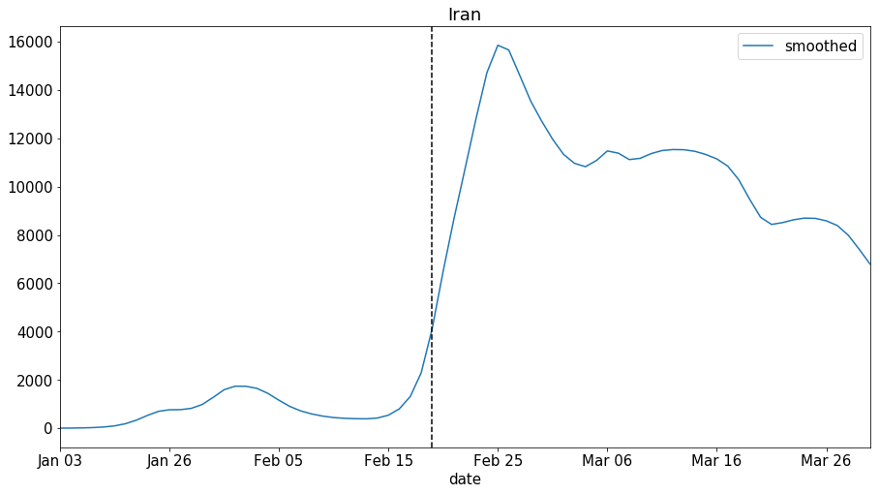}}
\caption{The Iranian case: Daily trends on velocity and acceleration of the \# of tweets (top) and divided phases detected by vertical dash lines (bottom).}
\label{fig:vel_acc_iran}
\end{figure}

\clearpage
\textbf{\\ \\}
\subsubsection{Vietnam}

\begin{figure}[h!]
\centerline{\includegraphics[width=1.35\linewidth]{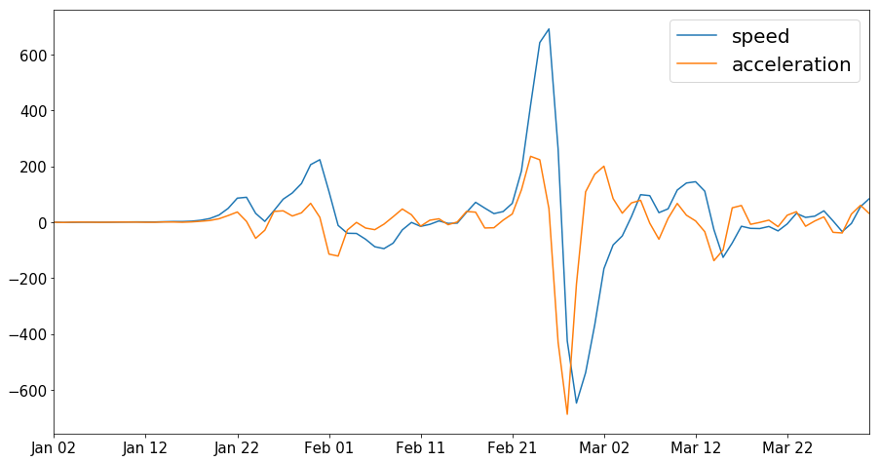}}
\textbf{\\}
\centerline{\includegraphics[width=1.35\linewidth]{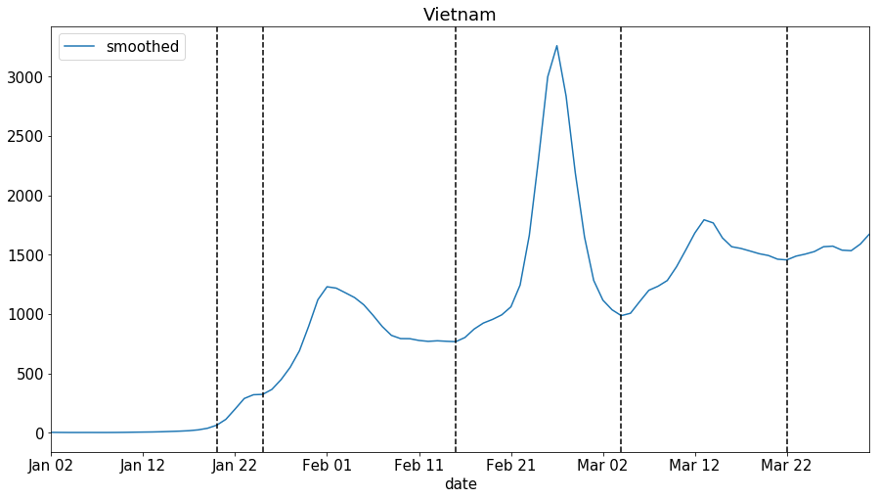}}
\caption{The Vietnamese case: Daily trends on velocity and acceleration of the \# of tweets (top) and divided phases detected by vertical dash lines (bottom).}
\label{fig:vel_acc_vietnam}
\end{figure}

\clearpage
\textbf{\\ \\}
\subsubsection{India}

\begin{figure}[h!]
\centerline{\includegraphics[width=1.35\linewidth]{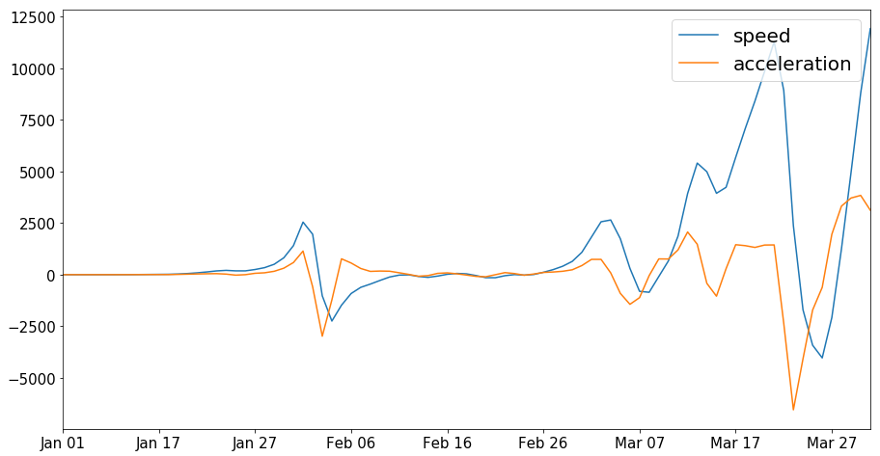}}
\textbf{\\}
\centerline{\includegraphics[width=1.35\linewidth]{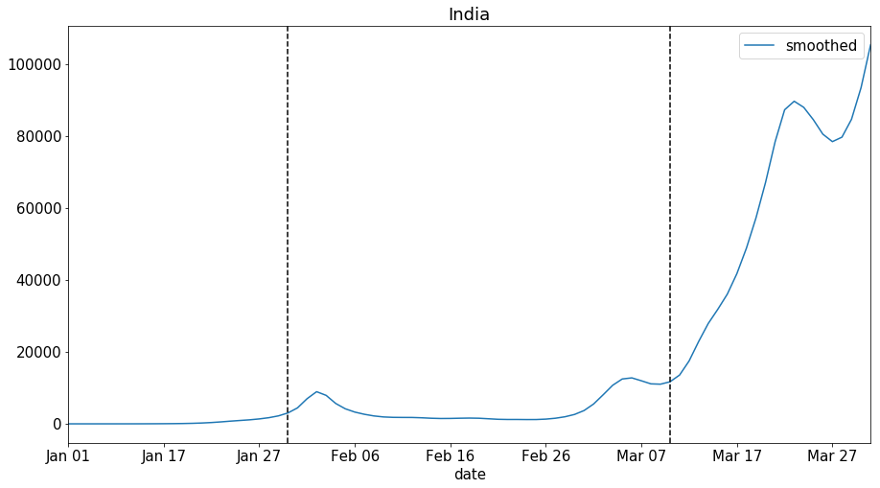}}
\caption{The Indian case: Daily trends on velocity and acceleration of the \# of tweets (top) and divided phases detected by vertical dash lines (bottom).}
\label{fig:vel_acc_india}
\end{figure}

\clearpage
\subsection{Appendix 2. Daily COVID-19 Confirmed Case and Tweet Count Trends by Country}

\subsubsection{Iran}
On February 19, two people tested positive for SARS-CoV-2 in the city of Qom\footnote{COVID-19 pandemic in Iran. Wikipedia 2020. \href{https://bit.ly/3ftQDV5}{https://bit.ly/3ftQDV5}.}. After this date, we see a significant surge in the number of tweets, and it reaches a peak in a few days (i.e., a peak shown on February 25). On February 23, the government changed the alert from white to yellow. Although the number of confirmed cases keeps increasing, the number of tweets starts to decrease gradually with a little fluctuation, as shown in Figure~\ref{fig:trends.iran}. Therefore, the trends of these two numbers show different patterns in contrast to Korean tweets. Meanwhile, the government gradually increased preventive measures, and several cities with the highest rate of infection were announced hot spots or red zones. Overall, they did not place the whole country under the red alert. However, the government announced new guidance and banned all trips on March 25. On March 28, the president said that 20\% of the country’s annual budget would be allocated to fight the virus, which might be implicitly a sign of the red alert.

\subsubsection{Vietnam}
On January 23, 2020, Vietnam officially confirmed the first two COVID-19 patients, who come from Wuhan, China\footnote{COVID-19 pandemic in Vietnam. Wikipedia 2020. \href{https://bit.ly/35BOyC2}{https://bit.ly/35BOyC2}.}. After that, the number of tweets increased sharply and reached to peak in early February, as shown in Figure~\ref{fig:trends.vietnam}. Although a few new cases were detected, the number of tweets tended to decrease and remained stable. In the second half of February, there are no new cases. However, the number of tweets increased rapidly and created a new peak. This peak could not remain for a long time. Two possible reasons can explain this trend. The first is that the pandemic has spread over the world. The second is that the last cases in Vietnam were treated successfully. After a long time with no new cases, Vietnam had confirmed continuously new cases in Hanoi and many other cities from March 6. The number of tweets of this phase increased again and remain stable at a relatively higher level than the initial phase.

\subsubsection{India}
The first case of COVID-19 was confirmed on January 30, 2020\footnote{COVID-19 pandemic in India. Wikipedia 2020. \href{https://bit.ly/37wIdsN}{https://bit.ly/37wIdsN}.}. The number of cases quickly rose to three on account of students returning from Wuhan, China. Throughout February, no new cases were reported, and the first weeks of March also saw a relatively low number of cases. The number of cases, however, picked up numbers from the fourth week of March, notable were the 14 confirmed cases of Italian tourists in the Rajasthan province. This eventually led to the government of India declaring a complete lock-down of the country. The daily number of tweets followed a similar trend as that of the number of cases as depicted in Figure~\ref{fig:trends.india}. First confirmed cases around January 30, 2020, caused a sudden spike in the number of tweets, that subsided in February. First COVID-19 fatality on March 12 and some other COVID-19 local events led to an exponential increase in tweets. The tweets peaked on March 22 when the government declared lock-down of areas with infected cases and started trending downwards after that. It is strange that the government's declaration of nationwide lock-down on March 24 only caused a small spike in the number of tweets and trend continued downwards. However, March 31 saw a significant spike in the number of tweets owing to confirmation of mass infections in a religious gathering. Overall, the tweet trends seem to be synonymous with the government's release of official information (e.g., \# confirmed cases and fatalities on COVID-19).

\clearpage
\subsection{Appendix 3. Daily Topical Trends Shown in Social Media by Country}
\subsubsection{South Korea}
Please refer to the ``Basic Daily Trends -- South Korea'' subsection in the manuscript for the detailed descriptions.

\begin{figure}[hbt!]
\centerline{\includegraphics[width=1.3\linewidth]{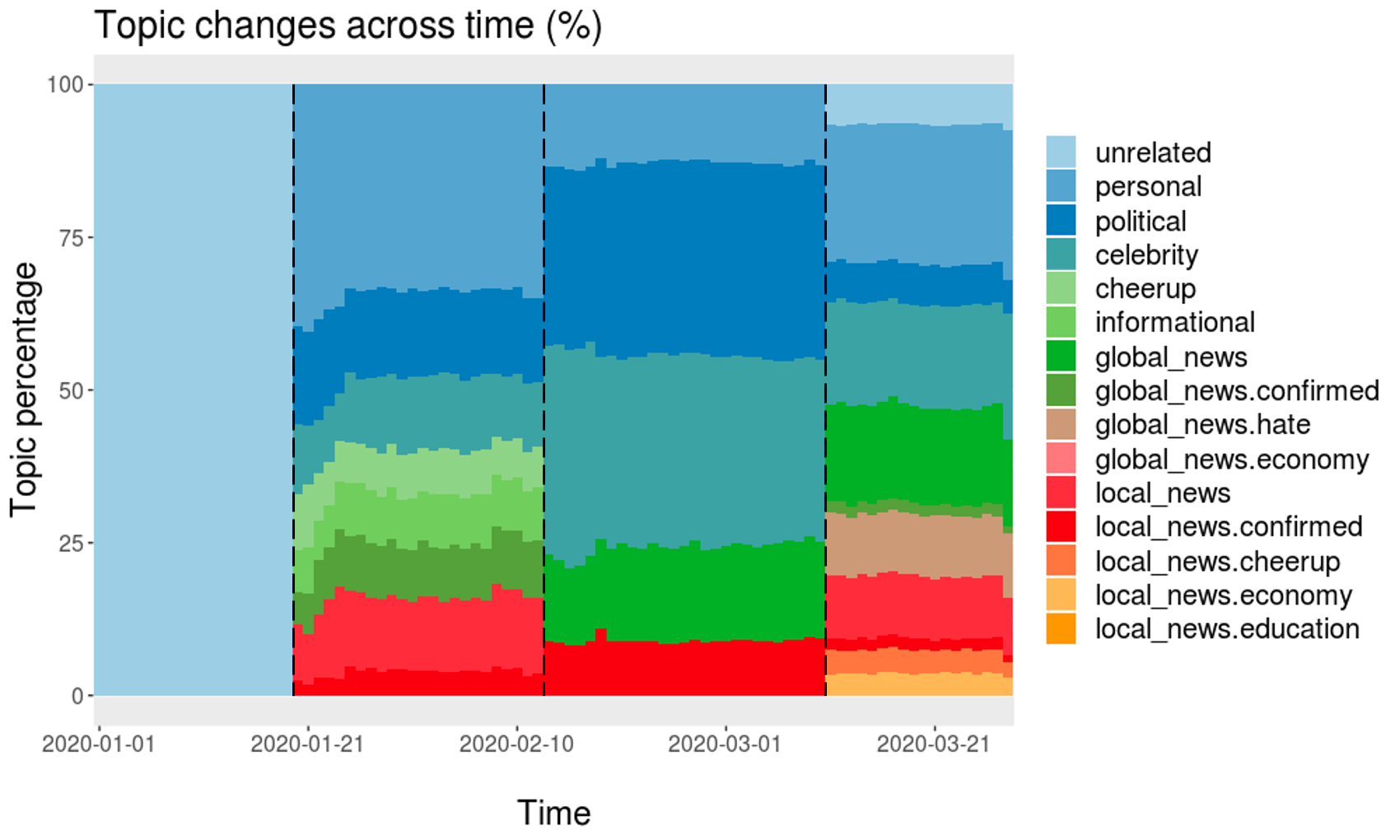}}
\centerline{\includegraphics[width=1.3\linewidth]{figure/topic_change_total_korea.png}}
\centerline{\includegraphics[width=1.3\linewidth]{figure/country_change_total_korea.png}}
\caption{Daily topical trends on South Korea: based on \% (top), based on \# of tweets (mid), based on \# of tweets country names mentioned (bottom).}
\label{fig:topic_change_korea_appen}
\end{figure}

\clearpage
\textbf{\\}
\subsubsection{Iran}
\begin{figure*}[hbt!]
\centerline{\includegraphics[width=0.75\linewidth]{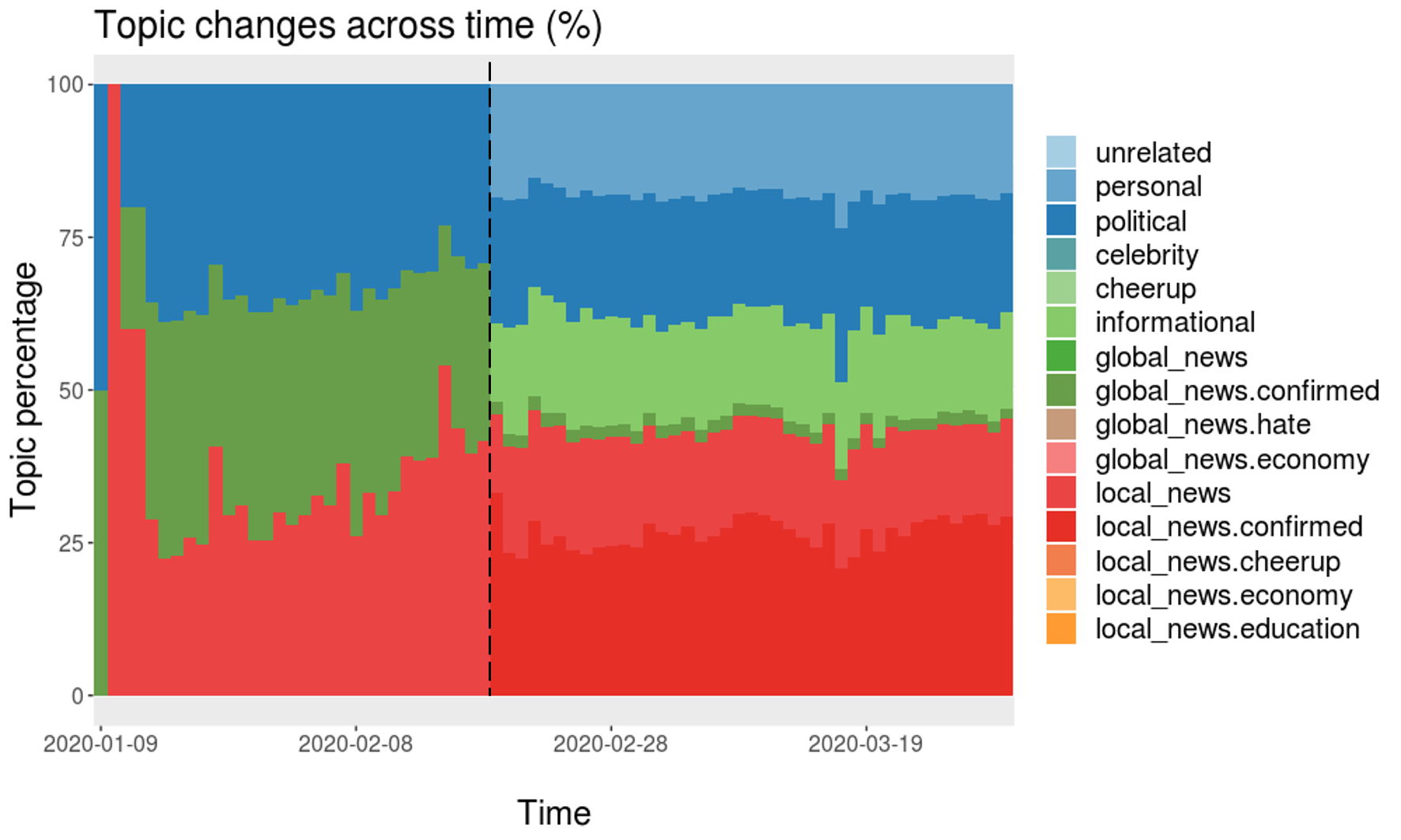}}
\centerline{\includegraphics[width=0.75\linewidth]{figure/topic_change_total_iran.png}}
\centerline{\includegraphics[width=0.75\linewidth]{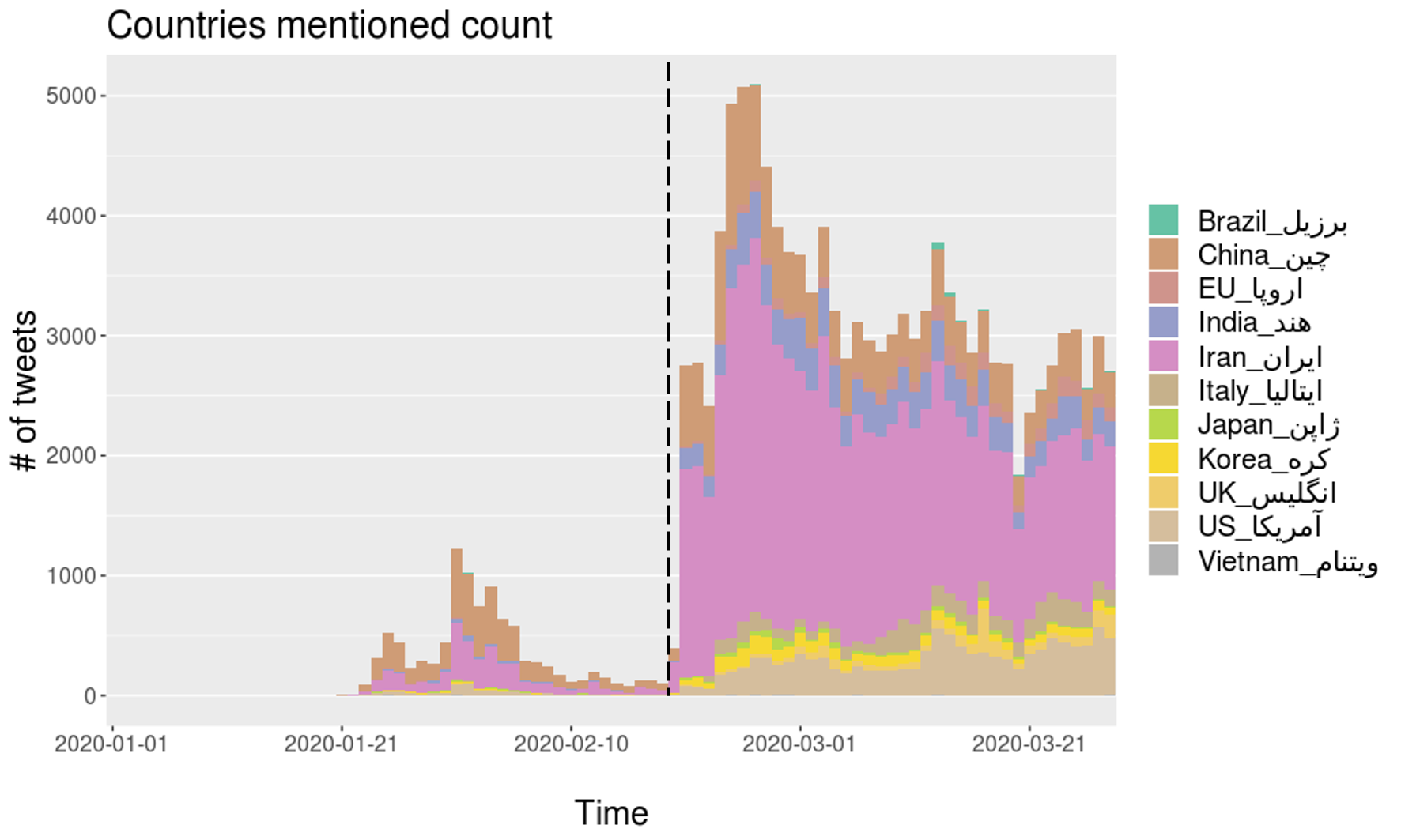}}
\caption{Daily topical trends on Iran: based on \% (top), based on \# of tweets (mid), based on \# of tweets country names mentioned (bottom).}
\label{fig:topic_change_iran_appen}
\end{figure*}

Figure~\ref{fig:topic_change_iran_appen} top and mid illustrate two topical phases, their proportions, and daily topical frequencies in Farsi tweets. Phase 0 includes global news about China and unconfirmed local news that reflects the fear of virus spread in the country. Political issues form a remarkable portion of tweets in this phase, as the country has been struggling with various internal and external conflicts in recent years, and there was a congressional election in Iran. In phase 1, a significant increase in the number of tweets occurs, where local news regarding the virus outbreak constitutes the majority. An intriguing finding is that informational tweets about preventive measurements overshadow global news, which can be explained by the sociology of disaster that when people in a less developed country are at risk, they naturally tend to share more information. However, political tweets are still widespread because of the reasons above and public dissatisfaction about the government response to the epidemic. This finding is also highlighted in Figure~\ref{fig:topic_change_iran_appen} bottom that the U.S. is the most mentioned name after Iran and China. One possible explanation is that the outbreak puts another strain on the frail relationship between Iran and the U.S.

\clearpage
\textbf{\\}
\subsubsection{Vietnam}
\begin{figure*}[hbt!]
\centerline{\includegraphics[width=0.75\linewidth]{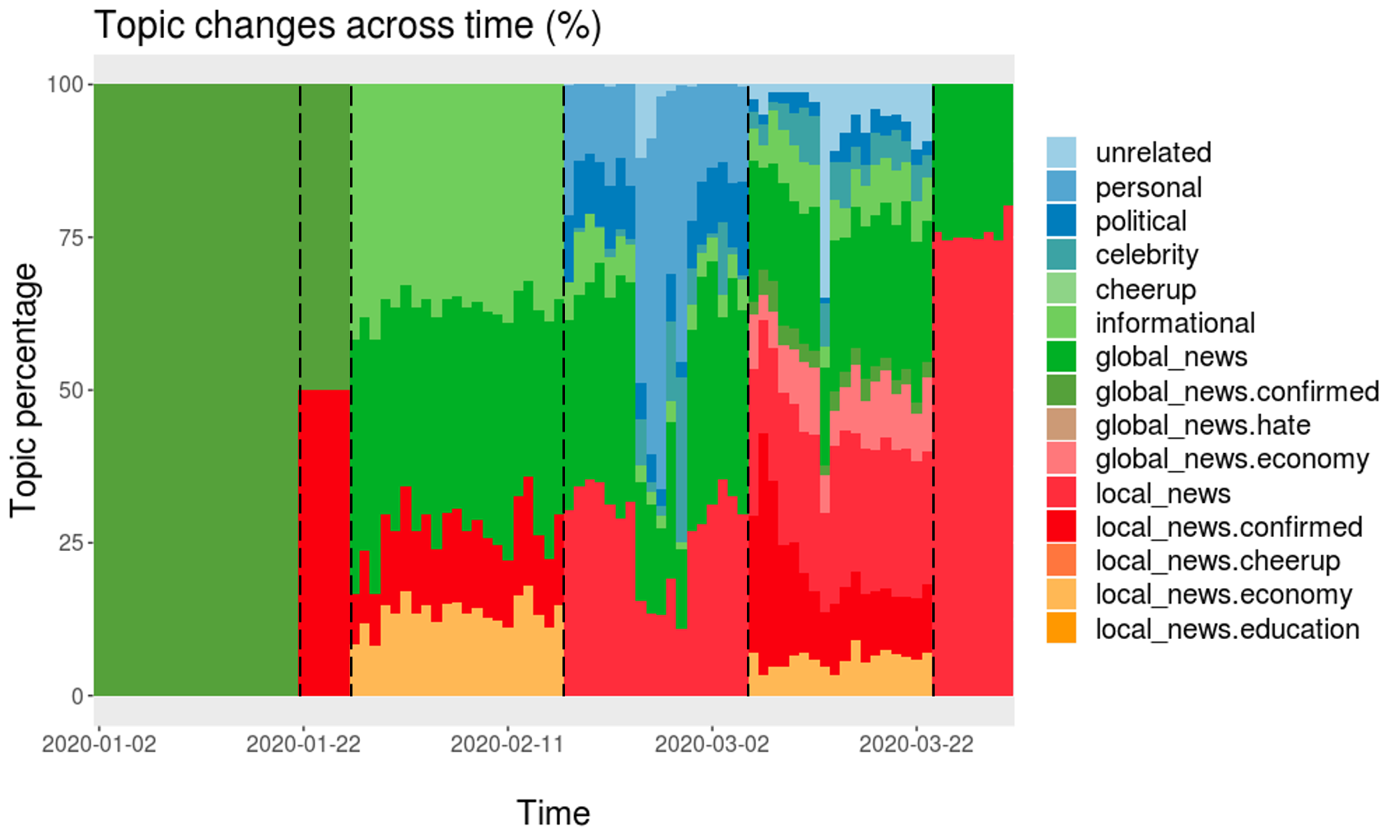}}
\centerline{\includegraphics[width=0.75\linewidth]{figure/topic_change_total_vietnam.png}}
\centerline{\includegraphics[width=0.75\linewidth]{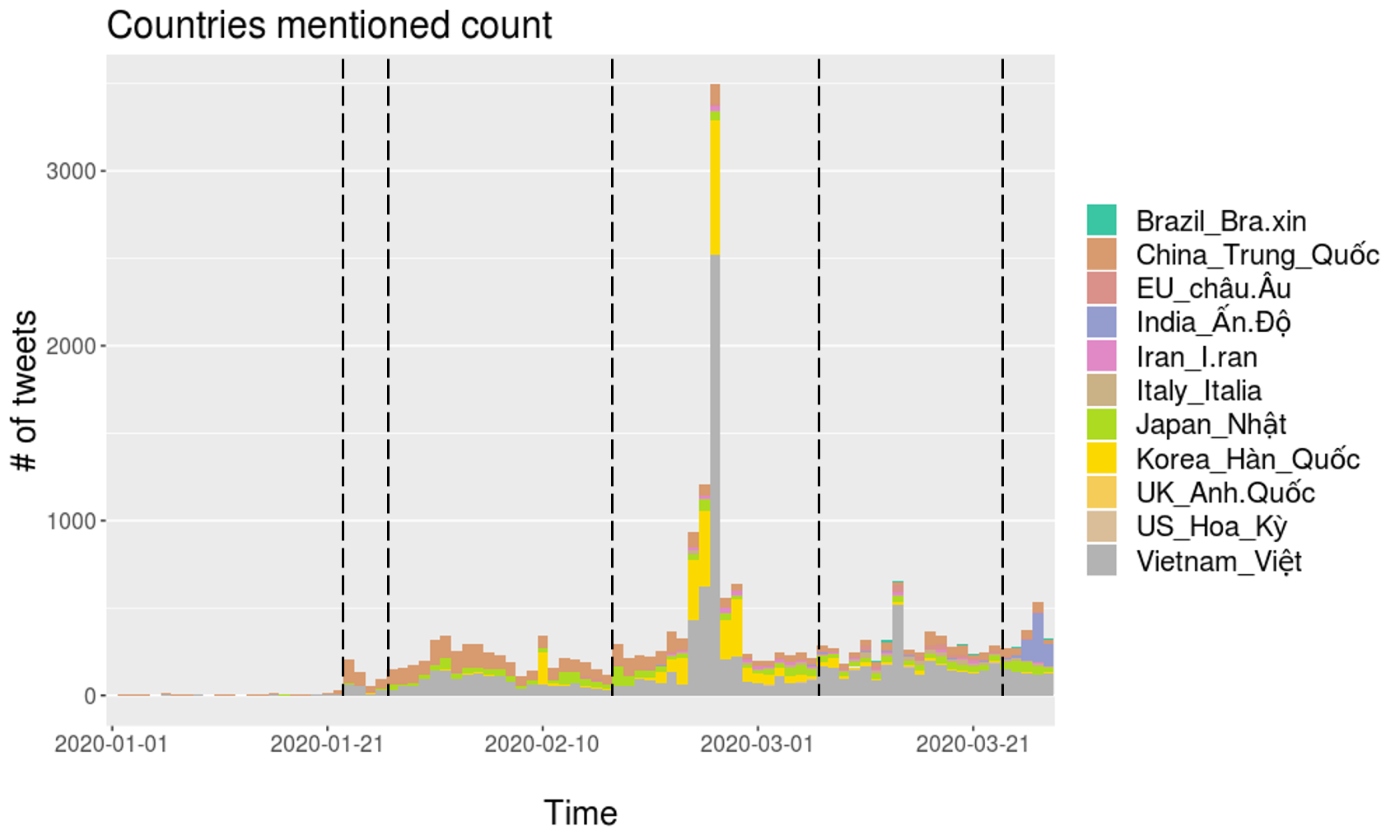}}
\caption{Daily topical trends on Vietnam: based on \% (top), based on \# of tweets (mid), based on \# of tweets country names mentioned (bottom).}
\label{fig:topic_change_vietnam_appen}
\end{figure*}

There are six topical phases with Vietnam, and they are visualized as in Figure~\ref{fig:topic_change_vietnam_appen} top and mid. Phase 0 related to global news because, in this period, Vietnam did not have any cases. From phase 1 to phase 5, topics diverged separately, but they focused on local news except phase 3. Phase 3 was the phase when no new cases in Vietnam were detected. We can see a common point of phase 0, and phase 3 is no new cases in Vietnam (local news), so tweets tended to talk more about global news. Especially in phase 3, we can see the increase of personal topics that most did not have in other phases. It was because a conflict event that related to Korean visitors made a huge of personal tweets. 
Next, we show the number of tweets that mentioned countries as in Figure~\ref{fig:topic_change_vietnam_appen} bottom. The most three countries mentioned are Vietnam, Korea, and China. Vietnam and China were mentioned frequently across phases because Vietnam is the local, and China is the original place of the pandemic. Besides, Korea was mentioned in many tweets, but they concentrated only on Phase 3. This is similar to topics changes due to the Korean visitor event in Vietnam.

\clearpage
\textbf{\\}
\subsubsection{India}
\begin{figure*}[hbt!]
\centerline{\includegraphics[width=0.75\linewidth]{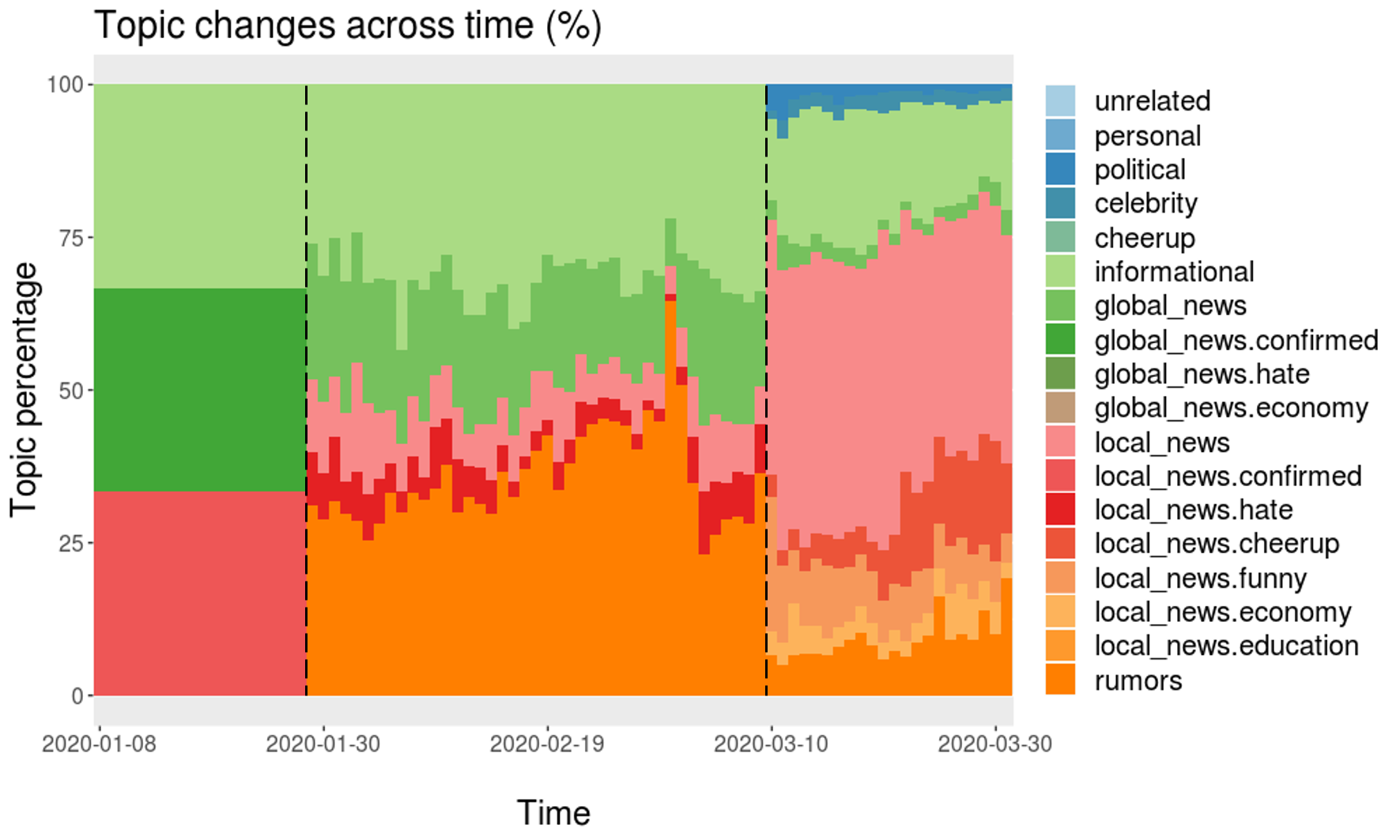}}
\centerline{\includegraphics[width=0.75\linewidth]{figure/topic_change_total_hindi.png}}
\centerline{\includegraphics[width=0.75\linewidth]{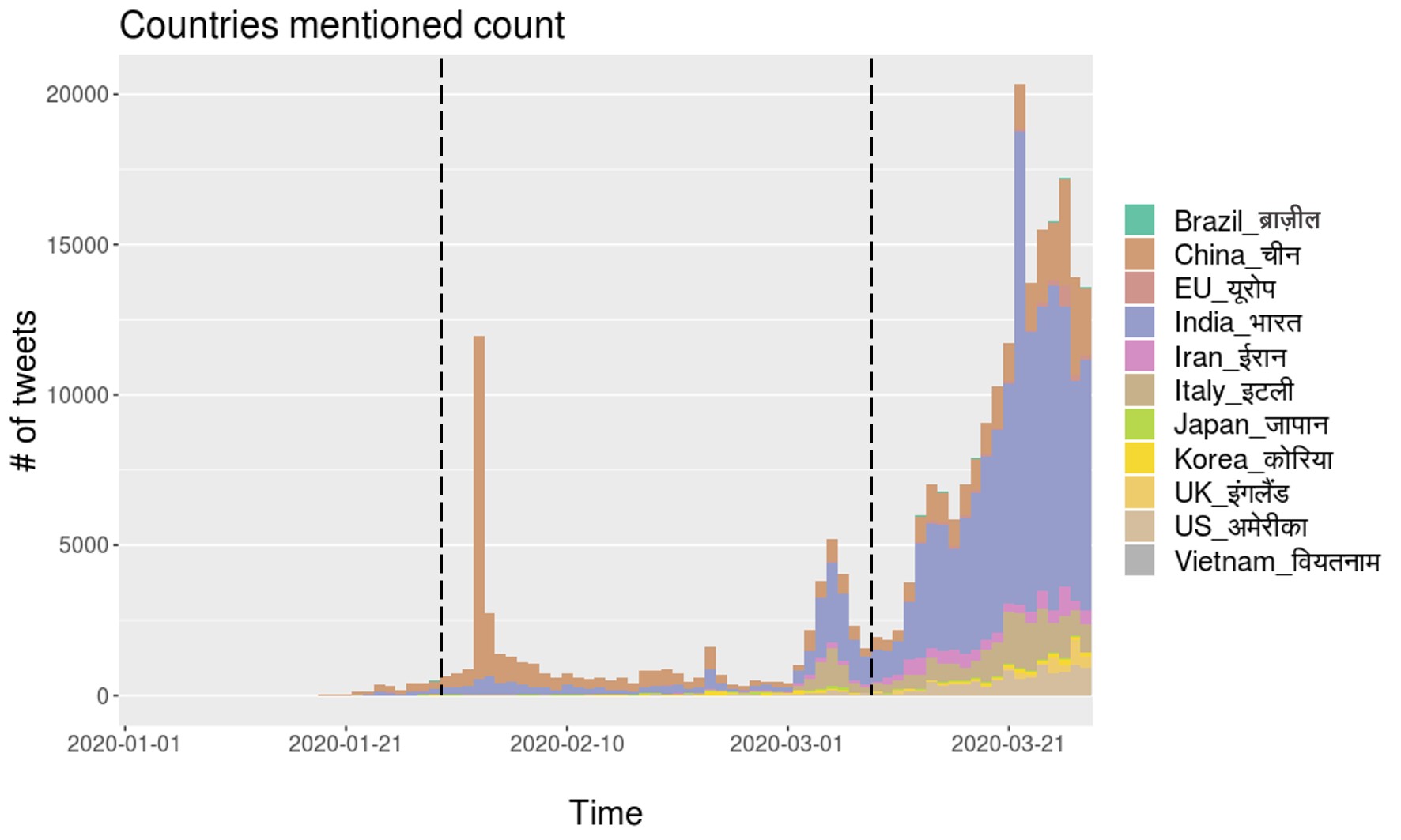}}
\caption{Daily topical trends on India: based on \% (top), based on \# of tweets (mid), based on \# of tweets country names mentioned (bottom).}
\label{fig:topic_change_india_appen}
\end{figure*}

We have established three topical phases for Hindi-written tweets (Figure~\ref{fig:topic_change_india_appen} top and mid). In case of India, the starting phase mainly has tweets that are focused on sharing information about COVID-19, and global news about COVID-19 in China. People tended to share the news about COVID-19 and useful information on how to be safe. Thereafter in phase 1, the number of topics become more diverse. Although the topics in this phase are related to information about the virus and global news, especially China, a large chunk of it is formed by rumors or misinformation. The number of tweets spike on January 30, 2020, when the first case was confirmed in India. Towards the end of Phase 1 is a spike in the number of tweets. This is primarily due to the beginning of announcements of some measures by the government to contain the virus (such as halting issuing new Visas to India). Lastly, in phase 2, a huge spike in the number of tweets is witnessed. The proportion of informational tweets decreases, whereas local news tweets confirming new cases increases. Regrettably, a marked portion of the tweets still consists of hateful content and misinformation. Interestingly enough, although the situation continued to worsen, tweets expressing dissatisfaction with the government are negligible.

Phase 3 also witnesses an increase in the mentions of other countries, especially Brazil and Europe, in addition to China and understandably, India as depicted in Figure~\ref{fig:topic_change_india_appen} bottom. This could be attributed to a growing number of the confirmed cases in Italy, Spain, and Brazil, as well as the news surrounding the use of Hydroxychloroquine in Brazil. The U.S. also finds considerable mentions due to the same reasons.

\end{document}